\documentclass[aps,prx,superscriptaddress,twocolumn]{revtex4-2}
\usepackage[utf8]{inputenc}
\usepackage[T1]{fontenc}

\usepackage{xcolor,graphicx,nicefrac}
\usepackage{amsmath}
\usepackage{amsfonts}
\usepackage{braket}
\usepackage{float}
\usepackage{hyperref}
\usepackage{xcolor}
\usepackage{epsfig}  
\usepackage{units}  
\usepackage{amsmath}
\usepackage{mathtools,amsfonts,amssymb,amsthm, bm, nccmath}
\usepackage{braket}
\usepackage{amsfonts} 
\usepackage{amssymb}
\usepackage{calrsfs}
\usepackage{ bbold }
\usepackage{ tipa }
\usepackage{amsmath,url}
\usepackage{enumitem}
\usepackage{eucal}
\usepackage{cancel}
\usepackage{algorithm}
\usepackage[noend]{algpseudocode}
\usepackage{cleveref}
\usepackage{ wasysym }
\usepackage{amsmath}
\usepackage{upgreek}
\usepackage[caption=false]{subfig}
\usepackage{mwe}

\newcommand{\order}[1]{\mathcal{O} \left( #1 \right)}

\newcommand{\bsection}[1]{\vspace{0.3cm}\begin{large}\textbf{#1} \end{large} \par}
\newcommand{\bbsection}[1]{\vspace{0.3cm}\textbf{#1.} }
\newcommand{\bbbsection}[1]{ \par \vspace{0.3cm} \noindent \begin{large}\textit{#1.} \end{large} }

\begin{document}

\title{Resource-efficient simulation of noisy quantum circuits and application to network-enabled QRAM optimization}

\author{Lu\'{i}s Bugalho}
\affiliation{Instituto Superior T\'{e}cnico, Universidade de Lisboa, Portugal}
\affiliation{Physics of Information and Quantum Technologies Group, Centro de F\'{i}sica e Engenharia de Materiais Avan\c{c}ados (CeFEMA), Portugal}
\affiliation{PQI -- Portuguese Quantum Institute, Portugal}
\affiliation{Sorbonne Université, CNRS, LIP6, 4 Place Jussieu, Paris F-75005, France}
\author{Emmanuel Zambrini Cruzeiro}%
\affiliation{Instituto de Telecomunica\c{c}\~{o}es, Lisbon, 1049-001, Portugal}

\author{Kevin C. Chen}
\affiliation{Research Laboratory of Electronics, Massachusetts Institute of Technology, Cambridge, MA 02139, USA}
\affiliation{Department of Electrical Engineering and Computer Science, Massachusetts Institute of Technology, Cambridge, MA 02139, USA}
\author{Wenhan Dai}
\affiliation{Department of Electrical Engineering and Computer Science, Massachusetts Institute of Technology, Cambridge, MA 02139, USA}
\affiliation{Department of Computer Science, University of Massachusetts, Amherst, Massachusetts 01003, USA}
\author{Dirk Englund}
\affiliation{Research Laboratory of Electronics, Massachusetts Institute of Technology, Cambridge, MA 02139, USA}
\affiliation{Department of Electrical Engineering and Computer Science, Massachusetts Institute of Technology, Cambridge, MA 02139, USA}

\author{Yasser Omar}
\affiliation{Instituto Superior T\'{e}cnico, Universidade de Lisboa, Portugal}
\affiliation{Physics of Information and Quantum Technologies Group, Centro de F\'{i}sica e Engenharia de Materiais Avan\c{c}ados (CeFEMA), Portugal}
\affiliation{PQI -- Portuguese Quantum Institute, Portugal}

\begin{abstract}
Giovannetti, Lloyd, and Maccone (2008)  proposed a quantum random access memory (QRAM) architecture to retrieve arbitrary superpositions of $N$ (quantum) memory cells via $O(\log(N))$ quantum switches and $O(\log(N))$ address qubits. Towards physical QRAM implementations, Chen \textit{et al.} (2021) recently showed that QRAM maps natively onto optically connected quantum networks with $O(\log(N))$ overhead and built-in error detection. However, modeling QRAM on large networks has been stymied by exponentially rising classical compute requirements. Here, we address this bottleneck by: (i) introducing a resource-efficient method for simulating large-scale noisy entanglement, allowing us to evaluate hundreds and even thousands of qubits under various noise channels; and (ii) analyzing Chen et al.'s network-based QRAM as an application at the scale of quantum data centers or near-term quantum internet; and (iii) introducing a modified network-based QRAM architecture to improve quantum fidelity and access rate. We conclude that network-based QRAM could be built with existing or near-term technologies leveraging photonic integrated circuits and atomic or atom-like quantum memories.
\end{abstract}

\maketitle
\bsection{Introduction}

A quantum random access memory (QRAM) is an essential computational primitive for many quantum algorithms. The ability to perform a QRAM query in $\log(N)$ time steps, where  $N=2^\mathrm{n}$  is the number of memory cells, implies polynomial speed-ups for applications such as quantum machine learning \cite{Biamonte2017}, matrix inversion \cite{Harrow2009}, quantum imaging \cite{Kiani2020}, and quantum searching \cite{Grover1996}. Despite its clear importance to quantum information processing, a QRAM has yet to be realized experimentally. Hence, finding a suitable architecture that can be realized in the near-future remains an active research subject in the theoretical and experimental domains.

In this article, we present a method to simulate large-scale entanglement accounting for various sources of noise. We are able to efficiently simulate circuits with thousands of qubits under dephasing, amplitude damping, and CNOT errors. Based on our simulation model, we present a QRAM architecture for photonic network-based QRAM based on Ref.~\cite{Chen2021}. The feasibility assessment is based on realistic parameters extracted from recent experiments, which we will refer to throughout the article.

A classical RAM~\cite{Jaeger1997} consists of a binary tree leading to a final layer of memory cells, each corresponding to an unique address. The address is represented as a series of bits, with each bit corresponding to a layer of the binary tree. Each bit of an address describes how the bus signal propagates in the layer: to the right or to the left child node. Hence, the nodes of the binary tree act as switches for the address. When provided with a $n$-bit address, the RAM returns a bit string $f_k$ associated to the memory cell labeled $k$. This is called the fan-out scheme~\cite{Giovannetti2008}.

A QRAM is the quantum analog of the RAM, similarly consisting of addresses, \textit{quantum} switches, and memory cells in the form of qubits. In particular, with a quantum address state, over the set of address qubits $a$, given by $|\psi_\text{in}'\rangle = \sum_{j=1}^\mathrm{n}\alpha_j|j\rangle_a$, one can retrieve data from a superposition of memory cells. A QRAM query is defined via the following transformation,
\begin{equation}
    |\psi_\text{in}\rangle = |\psi_\text{in}'\rangle|\emptyset\rangle_b \longrightarrow |\psi_\text{out}\rangle = \sum_{j=1}^N\alpha_j|j\rangle_a|D_j\rangle_b
    \label{eq:QRAMdef}
\end{equation}
where $|\emptyset\rangle$ represents an ancillary state, over the bus qubit $b$, which transforms into the retrieved data state after querying. In this article, we will restrict our investigations to classical data, i.e. $|D_j\rangle$ are separable bits. A direct conversion of classical fan-out protocol to the quantum realm is inefficient since it requires maintaining quantum coherence over an exponential number of connections~\cite{Giovannetti2008}.

Three main schemes have been investigated to date: the fan-out scheme that was already described, the bucket brigade model, and the teleportation-based scheme. Important figures of merit for the QRAM are the fidelity of the above transformation and the query time. For a detailed study and comparison of the first two schemes, please refer to Ref.~\cite{Hann2021a}. 

In the bucket brigade (BB) model~\cite{Giovannetti2008,Giovannetti2008a}, the number of qubits of the device scales as $O(2^n)$, as does the number of gates. Moreover, the original protocol ~\cite{Giovannetti2008} includes an additional third state in each node, called the ``wait'' state in order to prevent the exponential scaling of the amount of decoherence, with respect to the memory size. However, Hann \textit{et al.}~\cite{Hann2021a} have shown that the origin of the noise resilience of the BB model is the amount of entanglement among the memory's components and not the presence of the ``wait'' state, as one can devise a BB model without the ``wait'' state that still achieves a polynomial scaling of the decoherence with respect to the number of memory addresses $n$. 

More recently, Chen \textit{et al.} presented a photonic network-based QRAM scheme~\cite{Chen2021} that makes use of quantum teleportation of addresses from a quantum computer to the QRAM binary tree. Such a scheme greatly increases the protocol's efficiency by teleporting the registers to the layers (initially prepared in GHZ states) in parallel as opposed to in series, thereby circumventing the event of a single qubit loss collapsing the entire tree state. Additionally, the proposed QRAM maps onto quantum networks, leading to potential applications in distributed quantum computing and sensing.

However, Chen \textit{et al.} left as an open challenge the simulation of the scheme on large-scale networks since the computational complexity scales exponentially with the number of qubits. In this work, we bypass this problem resorting to more efficient ways of modeling the noise in stabilizer states. Moreover, this method generalizes to other quantum networking tasks with similar constructions, such as protocols for distributed quantum computation.

\begin{figure}[H]
    \centering
    \includegraphics[width=\columnwidth]{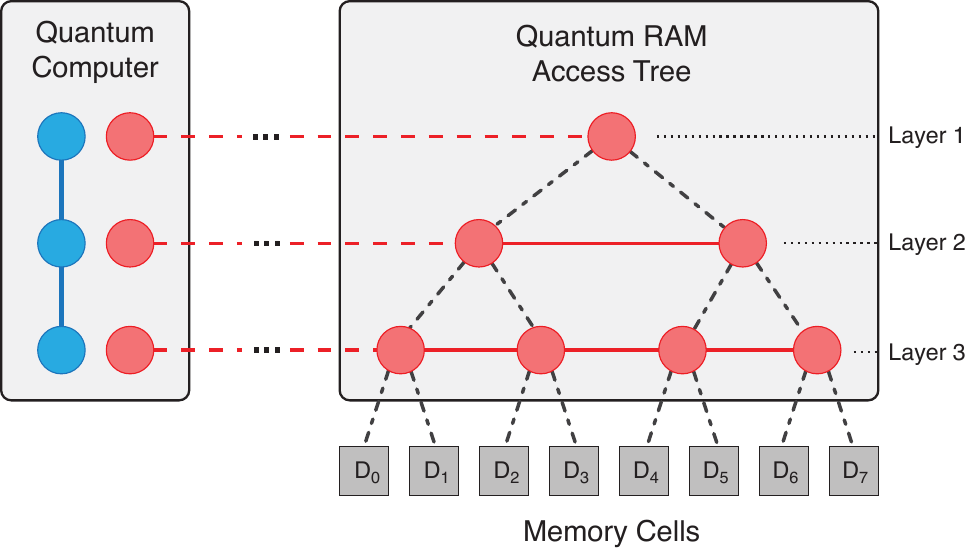}
    \caption{\textbf{Overview of a teleportation-based QRAM architecture.} A quantum RAM in the form of a binary tree comprises GHZ states for each physical layer. The left-most node of each layer $i$ is entangled with an ancillary qubit in a remote quantum computer, which hosts the query address qubits (blue). Bell state measurement in the quantum computer then teleports the address state onto the access tree. The elementary operations to constructing GHZ states in a photonic integrated circuit (PIC) QRAM are identical to the ones over \cite{Chen2021}.}
    \label{fig:QRAM_architecture}
\end{figure}

This comes in line with the fact that distributing entanglement is central in quantum information processing schemes ranging from quantum computing to sensing to communications~\cite{Toth2012a,Sidhu2019,Murta2020}. Simulation of distributed entanglement in a network setting, be it a long-distance network such as a possible future quantum internet~\cite{Wehner2018}, or small-distance quantum local area network (QLAN)~\cite{Alshowkan2021a}, is important to assessing the limitations imposed by near-term quantum technologies. The architecture of the QRAM considered in this paper, building on photonic network-based QRAM proposed in Ref.~\cite{Chen2021}, involves a series of exponentially growing GHZ states, with the largest having as many qubits as there are memory cells. Each GHZ state spans across a physical layer in the QRAM architecture, and the number of nodes per layer grows exponentially with the number of memory cells $2^n\equiv N$ to be addressed, as shown in Fig.~\ref{fig:QRAM_architecture}. 

Computer simulations of noisy quantum processes in such a system quickly becomes computationally intensive~\cite{VandenNest2011,Jozsa2014,Takahashi2020} due to the density matrices growing exponentially in size with the number of qubits.  Even though the entire QRAM protocol definition, $i.e$ the retrieval of data given an input address (see Eq.~\ref{eq:QRAMdef}), requires more than just Clifford operations, creating the routing state over the QRAM architecture only uses Clifford gates. These operations are the ones used to create these GHZ states and the teleporting the address state onto the QRAM access layers. Moreover, the operations required to access the QRAM after the routing state is distributed over the routing nodes only grows with the logarithm of the number of qubits of the QRAM, in comparison to the linear amount of operations required to create the routing state. This is the reason why noise in the system mostly comes from the GHZ states before access. In particular, this set of operations to create the routing state can be classically simulated efficiently~\cite{Jozsa2014}. This approach enables an explicit and efficient description of all the intermediary states, up to local unitary corrections. In this article, we develop efficient methods to simulate large-scale noisy entanglement by characterizing the impact of noise at all intermediate steps, and apply these tools to simulate a noisy QRAM.

There are several architectures for a QRAM. Here, we focus on the optically mediated quantum network-based QRAM architecture introduced in Ref.~\cite{Chen2021}, as it offers several key benefits: implementation in quantum networks compatible with envisioned quantum internet architecture and quantum data centers, and faster query times and possibility of executing in a non-local manner by means of teleportation. Hence, this scheme works under any network-like architecture, be it locally ($e.g.$ on a chip) or across large distances ($e.g.$ over a quantum internet). Without loss of generality, we characterize each node of the architecture as one of a spin-photon network that could be implemented in photonic integrated circuits (PIC). 

The architecture of the QRAM is similar to previous models, such as BB and the fan-out models. The main difference concerns the execution of the protocol and the resources available at each node. In this architecture, one considers two agents: the quantum computer, which prepares the addresses, and the QRAM or quantum access tree (see Fig.~\ref{fig:QRAM_architecture}). The quantum computer must provide an address state with $n=\log_2 N$ qubits, where $N$ is the total number of memories (for simplicity assume $n\in\mathbb{N}$). The QRAM has a binary tree architecture, with $n$ physical layers, where the $k$th layer ($k \in \{1,...,n-1\}$) has $2^{k-1}$ quantum nodes. As we describe next, in each physical layer, all the nodes share a GHZ state, which is used to teleport the address state onto the QRAM itself, allowing for an ancilla qubit to access the memories in the correct superposition.

\begin{figure}[t]
	\centering
	\includegraphics[width=\columnwidth]{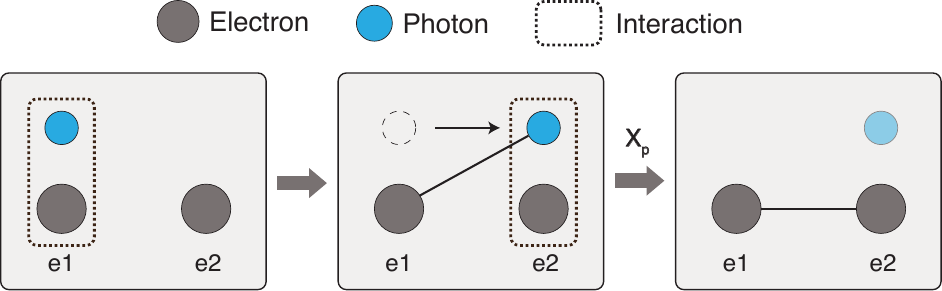}
	\caption{\textbf{Probabilistic CNOT.} Execution of a CNOT gate between two electrons, e1 and e2, mediated by a photon.}
	\label{fig:cnotmediated}
\end{figure}

As for the type of physical implementation chosen, and without loss of generality, we focus on a QRAM implementation involving solid-state spin qubits integrated into PICs, an approach that is promising in terms of scalability. In particular, we consider diamond nanophotonic cavities coupled with silicon-vacancy centers~\cite{Bhaskar2020,Wan2020} as each QRAM tree node. Each emitter contains an electronic spin that directly interacts with the photonic address register qubits and an accompanying nuclear spin acting as a long-lived memory. By entangling the electronic spin with the photon via cavity reflection, consecutive reflection of a photon off two neighboring nodes and subsequent heralding achieves spin-spin entanglement. This remote entangling strategy is repeatedly used to generate a GHZ state across each layer. Such operations are probabilistic (see Fig.~\ref{fig:cnotmediated}): the photon has a non-zero probability of being lost to the environment before reflecting off two cavities and arriving at the detector. On the other hand, it is possible to perform close to deterministic two-qubit gates between the electronic and nuclear spin qubits, albeit with a larger error~\cite{Nguyen2019,Bradley2021}. For this reason, we term this architecture teleportation-based deterministic QRAM, or TD-QRAM.

In these types of systems, the main contributors to errors are (i) spin phase errors (at rate $1/T_2$), (ii) spin flip errors (at rate $1/T_1$), and (iii) errors in hyperfine gates between electron and nuclear spins (see Supplementary Table I). We leave out photon-electron interactions, as one could conceive trading-off the efficiency $\eta$ for arbitrarily high fidelity in the cavity-reflection based scheme proposed in Ref.~\cite{Chen2021a} in the high-cooperativity and over-coupling regime.

Hence, we explore different values for $T_1$, $T_2$ of both electronic and nuclear spin qubits, and $p_\mathrm{e}$ and $p_\mathrm{n}$ for the probabilities of error in electronic and nuclear spin CNOTs. For the remaining of this article, we set $T_1^\mathrm{n} = 100 \ T_1^\mathrm{e} \equiv 100 \ T_1$ and $T_2^\mathrm{n} = 100 \ T_2^\mathrm{e} \equiv 100 \ T_2$. Nuclear spins have a higher coherence time as they are much less coupled to the noisy spin-bath compared to electronic spins. Reported values of characteristic times go, experimentally,  up to $T_1^\mathrm{e} \sim 1~\mathrm{s}$, $T_2^\mathrm{e} \sim 10~\mathrm{ms}$~\cite{Sukachev2017}, and there are theoretical predictions of being able to reach   $p_\mathrm{e}, p_\mathrm{n} = 10^{-2} \sim 10^{-4}$~\cite{Duan2004a,Calderon-Vargas2019}. Moreover, we detail other important physical parameters of this type of system, used for the simulations, in Supplementary Table I.

\begin{figure*}[t]
    \centering
    \includegraphics[width=\textwidth]{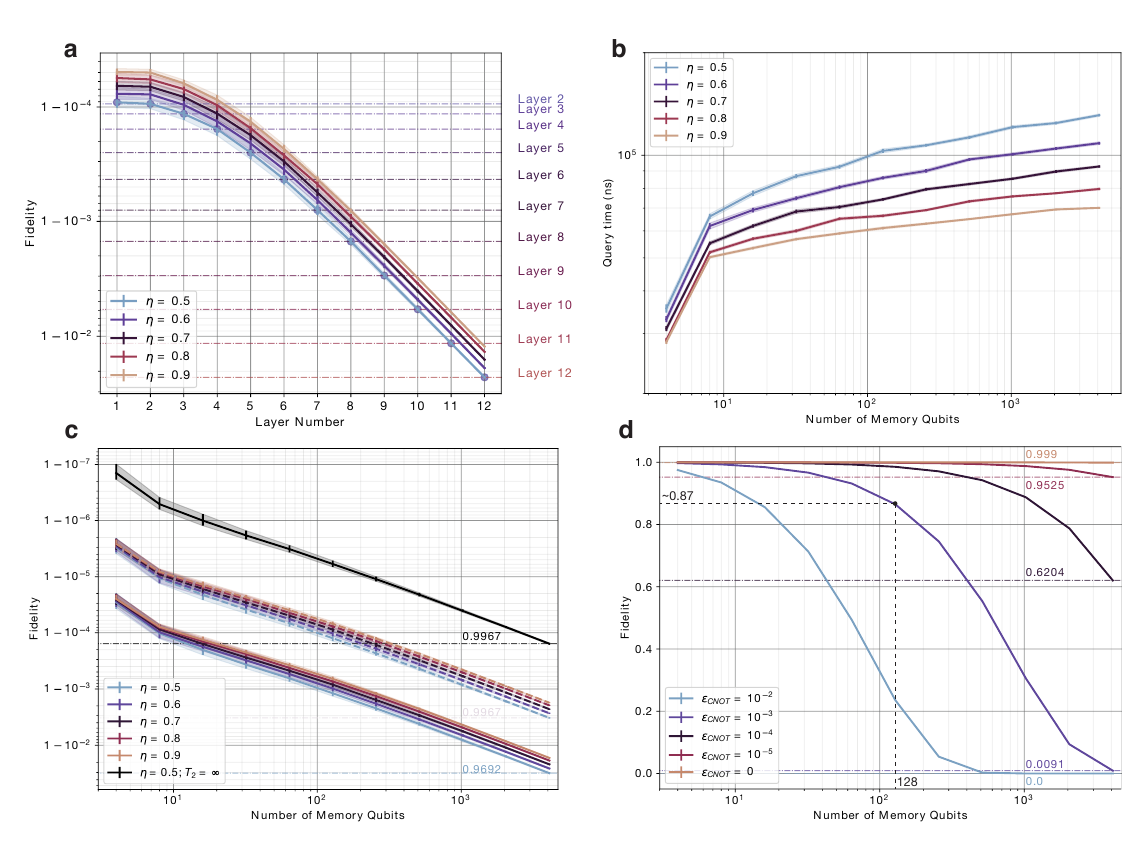}
    \caption{\textbf{TD-QRAM Simulations.} (a) TD-QRAM access protocol for 12 layers, with the efficiency of generating a Bell pair swept from $\eta = 50\%$ to $\eta = 90\%$. The noise analysis considers only dephasing and damping errors. The final fidelity is calculated according to Eq.~\ref{eq:QRAMfidelity}, with $T_1=20~\mathrm{ms}$, $T_2=10~\mathrm{ms}$, and $\epsilon_{\text{CNOT}} = 0$ for each layer. (b) Query times with varying sizes from 2 layers to 12 layers, and sweeping the efficiency of generating a Bell pair from $\eta = 50\%$ to $\eta = 90\%$. There is an expected logarithmic scaling of the query time with the number of qubits. (c) TD-QRAM noise analysis with dephasing errors, $T_2 = 10~\mathrm{ms}$ (filled lines) and $T_2 = 100~\mathrm{ms}$ (traced lines), with fixed amplitude-damping error $T_1 = 2~\mathrm{s}$. We consider different QRAM sizes from 2 layers to 12 layers as well as various efficiencies of generating a Bell pair from $\eta = 50\%$ to $\eta = 90\%$. (d) TD-QRAM noise analysis with noisy CNOTs, $p_\mathrm{e} = p_\mathrm{n} \in \{ 0, 10^{-5}, 10^{-4}, 10^{-3}, 10^{-2} \}$, for a QRAM with the number of layers ranging from 2 to 12. The dephasing time is fixed at $T_2 = 100~\mathrm{ms}$, and the amplitude-damping time is fixed at $T_1 = 2~\mathrm{s}$. The efficiency of generating a Bell pair is fixed at $\eta = 90\%$. The final fidelity mainly depends on the number of noisy CNOTs performed throughout the protocol and has little dependence on the efficiency. All the error bars over the data correspond to the error of the average value over 100 simulations of the protocol.}
    \label{fig:QRAMlayers} \label{fig:twosteptimes} \label{fig:dephasing} \label{fig:dampdephcnots}
\end{figure*}

\bsection{Results}
\bbsection{Simulating the effects of decoherence for a TD-QRAM}
To simulate the QRAM initialization protocol, we use NetSquid~\cite{Coopmans2022} under the stabiliser formalism and extract all the parameters of the noise channels before implementing them in simulations, for instances: timing parameters for every qubit used throughout the simulation, all the noisy CNOTs with corresponding error probabilities, and to which qubits and at which step it is applied. From here, we compute the fidelity of the final QRAM state by substituting all these values into the expressions presented in the methods.

We start by presenting the simulation of a $2^{12}$-qubit QRAM in Fig.~\ref{fig:QRAMlayers}. Here, we detail individually the fidelities of the GHZ state distributed at each physical layer of the QRAM. The fidelity of the full state of the QRAM is given by:
\begin{equation}
\begin{aligned} \label{eq:QRAMfidelity}
F(\text{QRAM}) &= \prod_{i=1}^{n-1} \mathcal{F}\left(\text{Layer}_{i},\ket{\text{GHZ}}_{2^{i-1}+1}\right), \\
    &\text{where }\ket{\text{GHZ}}_{q} = \frac{1}{\sqrt{2}} \left( \ket{0}^{\otimes q} + \ket{1}^{\otimes q} \right)
\end{aligned}
\end{equation}
$i.e.$, the fidelity of the entire tree (or the QRAM) is defined as the product of the fidelities of each physical layer (see Supplementary Methods for more details). We distinguish access fidelity from tree fidelity, where the former refers to the fidelity of the state retrieved after accessing the memory cells ($\ket{\psi_\text{out}}$ in Eq.~\ref{eq:QRAMdef}), and the latter refers to the multipartite state fidelity of the binary tree constituting the QRAM. Only the access fidelity depends on the address and bus qubits.

One observes an exponential decrease of the fidelity with the number of the layer (notice the logarithmic scaling on the $y$-axis corresponding to the fidelity). This agrees with the GHZ state size increasing exponentially with the number of layers, i.e. scaling $2^k$. When one qubit in this multipartite state suffers an error, the entire state is affected.

One critical figure of merit that we extract from the NetSquid simulations is the query time. As demonstrated in Ref.~\cite{Chen2021}, the query efficiency scales logarithmically with the number of qubits. Extracting from multiple queries of the QRAM, we obtain the query times (apart from a logarithmic factor derived from making the bus qubit traverse the binary tree) in Fig.~\ref{fig:twosteptimes}.

\bbsection{Dephasing and Damping Errors for TD-QRAM}
Considering only the effects of dephasing and amplitude-damping errors in the spin qubits, we take $T_2 = \infty~\mathrm{ms}$ for amplitude-damping errors only, and then $T_2 = 10~\mathrm{ms}$ and $T_2 = 100~\mathrm{ms}$ with a fixed $T_1 = 2~\mathrm{s}$~\cite{Sukachev2017}, see Supplementary Table I. We also set the CNOT error rate to 0. We present the simulation results for the TD-QRAM scheme under memory dephasing for increasing QRAM size, as shown in Fig.~\ref{fig:dephasing}.

Looking closely at Figure \ref{fig:dephasing}(c), one can observe that the effect of amplitude-damping shows an identical behavior to the one of dephasing and amplitude-damping combined, $i.e.,$ with the same type of scaling. However, it is residual comparatively to the effect of dephasing. This is easily explainable by the time-scales of the coherence times of the corresponding noises ($T_1$ and $T_2$) in the memory differ by orders of magnitude, with the first, $T_1$, being usually much longer than the latter, $T_2$, \textit{i.e.} $T_1\sim 1~\mathrm{s}$~\cite{Sukachev2017}. For this reason, its impact can be neglected relative to other sources of error.

\bbsection{Dephasing, Damping and Noisy CNOTs for TD-QRAM}
The only type of error missing in the analysis is the error derived from the use of noisy CNOTs. Illustrated in Fig.~\ref{fig:dampdephcnots}, the dephasing and damping errors minimally contribute to infidelity. We now analyse the case for noisy CNOTs on top of fixed $T_1 = 2~\mathrm{s}$ and $T_2 = 100~\mathrm{ms}$ (note we now switch to linear scale in the y-axis for the fidelity, due to the set of values present for the different simulations). For simplicity, we consider equal CNOT error probability, $\epsilon_\text{CNOT}$, for both \textit{electronic} and \textit{nuclear} CNOTs, and vary $\epsilon_\text{CNOT}$ from $10^{-5}$ to $10^{-2}$ as shown in Fig.~\ref{fig:dampdephcnots}:

These simulations show that the CNOT gates dominate the overall error in the QRAM state fidelity in the TD-QRAM. For instance, to access a 128-qubit QRAM, one needs fidelities of the CNOT gates to be somewhere near $99.9\%$ to obtain an access fidelity exceeding $90\%$. In this architecture, while the query times do not increase linearly with the size of the memory, the errors \textit{do}. Expectedly, applying an error to a single qubit of a GHZ state contributes in the same order for the entire state. 

The price to pay for performing CNOTs with such large error rates deterministically could be circumvented by near-perfect yet probabilistic CNOTs~\cite{Duan2004,Chen2021a} via cavity-based electron spin-photon interactions, as opposed to deterministic yet error-prone nuclear-electron spin coupling. In light of this, we explore a \textit{hybrid} teleportation-based QRAM architecture in the following section.

\bbsection{Teleportation-based Stochastic QRAM \label{sec:hybrid}}
In the TD-QRAM protocol, the entanglement generation and swap (Fig.~\ref{fig:block1}) operation are still probabilistic given the finite chance of photon loss. Hence, these \textit{probabilistic} CNOTs are done in parallel throughout each physical layer to improve efficiency. After an EPR pair is created between two electron spins, however, transferring entanglement onto the nuclear spins is a deterministic procedure. Thereby, the query time grows sub-linearly. As noted before addressing the TD-QRAM scheme, this \textit{deterministic} CNOT based on nuclear-electron spin interaction mainly dominates the infidelity of the GHZ state, motivating us to contemplate an alternative solution.

Since the decoherence errors from $T_1$ and $T_2$ contribute much less to the infidelity relative to electron-nuclear spin CNOT, replacing some of the noisy deterministic CNOTs with probabilistic CNOTs helps improve the fidelity despite reducing efficiency. As we will show, this leads to higher QRAM tree state fidelities, albeit with longer query times. We call this architecture `teleportation-based \textit{stochastic} QRAM', or TS-QRAM.

Relying solely on probabilistic CNOTs in every step of the protocol would be \textit{very} inefficient since the probability of generating a GHZ state diminishes exponentially with the number of nodes. In other words, if one entanglement attempt fails during construction of a GHZ state, the entire state collapses. Since each linking process is heralded, there are ways to circumvent this by choosing a specified order to perform the CNOTs, similar to entanglement swapping in a repeater chain~\cite{Coopmans2020e,Dai2020}. Here, the probabilistic swapping operations are equivalent to the probabilistic CNOTs, and measuring the middle node is analogous to joining smaller GHZ states to form a larger GHZ state. Abstractly, they describe the same problem, which allows us to use the solutions provided by Ref.~\cite{Dai2020}. Next, we present an in-depth analysis of the trade-off between fidelity and query rate as a function of error rates and physical implements.

\bbsection{Increasing $T_1$ and $T_2$}
To decrease the number of employed deterministic CNOTs, and taking into account that these always happen when the electronic spins interact with the nuclear spins, it is natural to consider dropping the nuclear spins altogether. This is motivated by the fact that we can perform CNOTs, albeit probabilistically, between the electron spins. The downside is that electron spins suffer from having shorter coherence times than their nuclear counterparts. Still, it is advantageous to consider such schemes to avoid the use of noisier deterministic CNOTs.

To minimize the consequently increased decoherence, one could conceive schemes for increasing the $T_1$ and $T_2$ times for the electrons, since these are the ones now causing the fidelity bottleneck, together with the required time to query the memory.

Presently, the SiV's electronic spin's $T_1$ time is shown to be longer than 1~s~\cite{Sukachev2017}, thereby posing no concern over depolarisation. On the other hand, its $T_2$ coherence time is limited to tens of milliseconds~\cite{Sukachev2017} even under dynamical decoupling. The main dephasing mechanism is attributed to the surrounding nuclear spin bath, which is weakly coupled to the electronic spin of interest via hyperfine interaction~\cite{Childress2006}. A potential avenue to improving the electronic spin's $T_2$ is therefore to ``purify'' its environment by materials engineering~\cite{Findler2020}. By producing SiV in a carbon-13 free matrix, for example, the coherence time may be further extended.

Nevertheless, our numerical analyses of the hybrid scheme show fidelities still exceeding $60\%$ for a reasonable CNOT error rate of $10^{-3}$ and 1024 memory cells, using a $T_2$ of $100~\mathrm{ms}$. For such a result, a probability of success of about $70\%$ for the CNOT is required.

\bbsection{The Teleportation-based Stochastic QRAM Protocol}
In the TD-QRAM protocol, there are two steps occurring in parallel across each layer in the QRAM:  one for generating EPR pairs across every other node and another for linking all the states into a larger GHZ state, via sharing EPR pairs in-between nodes holding the previously shared EPR pairs (see Fig.~\ref{fig:twosteps}).  This could be made in parallel because the linking operations are deterministic.

\begin{figure}[H]
	\centering
	\includegraphics[width=\columnwidth]{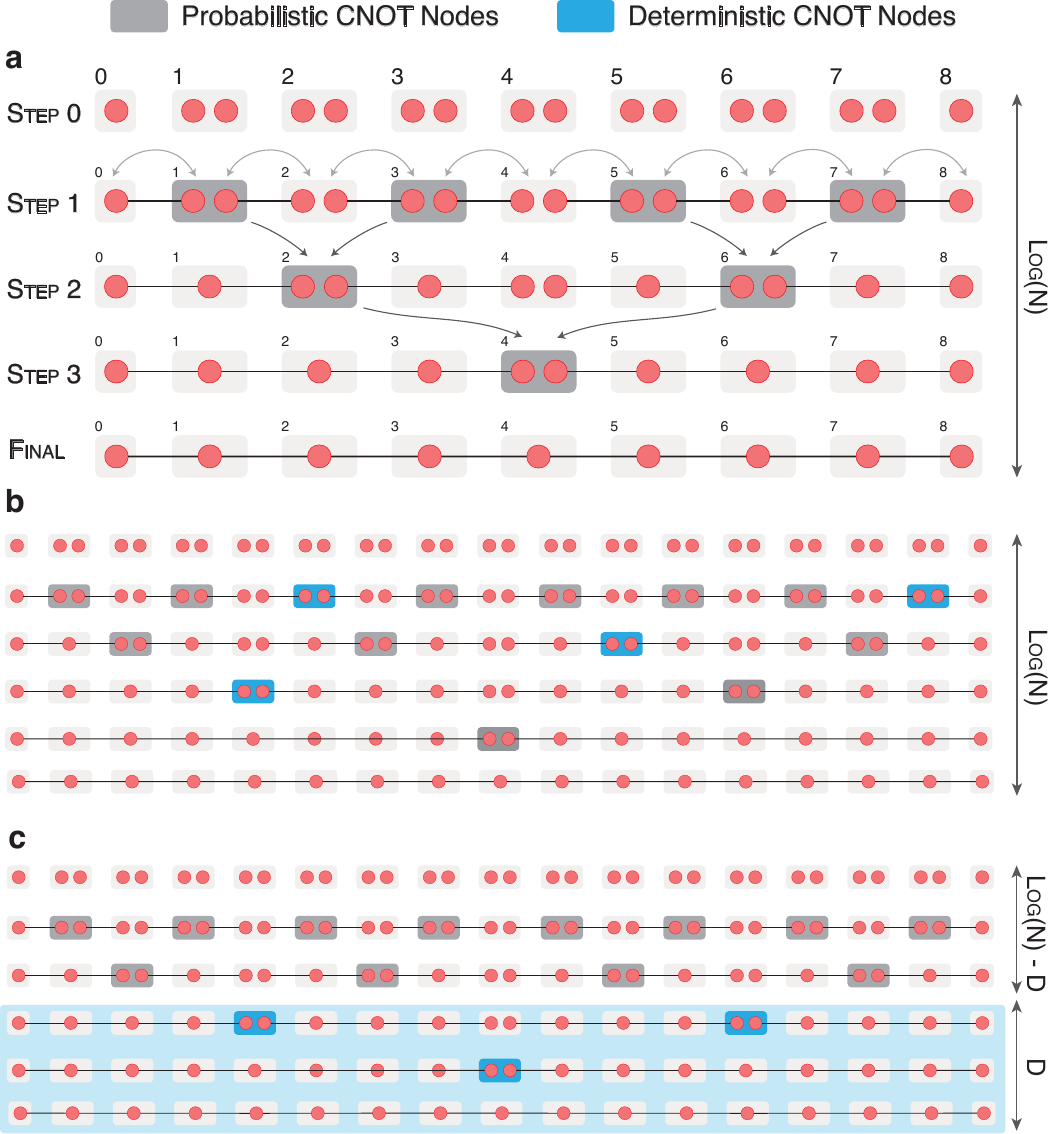}
	\caption{\textbf{Binary-tree-like approach of linking nodes and possible placement of deterministic CNOTs.} (a) The arrows represent heralding signals for the subsequent step, and the dark nodes represent the selected nodes for attempting entanglement at each time step. (b) Randomly distributed deterministic nodes across the $\log{N}$ distribution layers. (c) Intuitively distributed deterministic nodes with $D$ deterministic distribution layers. }
	\label{fig:linkorder} \label{fig:linkorderoptions}
\end{figure}

In the TS-QRAM protocol, however, we must now consider an order for the linking step that depends on the node's position, similar to the quantum repeater chain problem~\cite{Brand2020,Dai2020,Coopmans2022}. If a linking process fails, the subset of qubits that would have become entangled must be reset. The optimal strategy is then performing the linking process in a binary-tree-like approach~\cite{Dai2020}. This binary-tree order for the linking processes means that now a heralding signal for a successful link must be exchanged within the tree. Each parenting node will have two children, the right-child and the left-child. Each node only attempts entanglement if it receives heralding signals from both children nodes that have been successfully entangled themselves. Fig.~\ref{fig:linkorder} illustrates this procedure and defines the order.

Moreover, as mentioned before, the advantage of the TS-QRAM protocol is that probabilistic CNOTs are used to minimize state infidelity. One might consider the optimal placements for the deterministic CNOTs to maximize the GHZ state fidelity across each physical layer. We further introduce having an additional \textit{distribution} layer. This is the layer of the order binary-tree at which a linking step is attempted, as shown in Fig.~\ref{fig:linkorder}. These abstract layers are only needed to describe the order of the linking steps and help illustrate the optimal placements for the deterministic CNOTs.

For this reason, we present two possible options to solve the placement of deterministic nodes problem: the first is randomly choosing a set of nodes to be deterministic, regardless of their distribution layer. The second option is choosing the nodes that attempt to link entanglement at the higher steps, since if those attempts are unsuccessful, they take the biggest toll on the protocol requiring re-attempting every preceded step. We illustrate these two possible options in Fig.~\ref{fig:linkorderoptions}.

As we will verify later, we need a much smaller number of deterministic nodes if we place them in higher-level distribution layers. We first present simulation results for both cases.

\begin{figure*}[t]
    \centering
    \includegraphics[width=\textwidth]{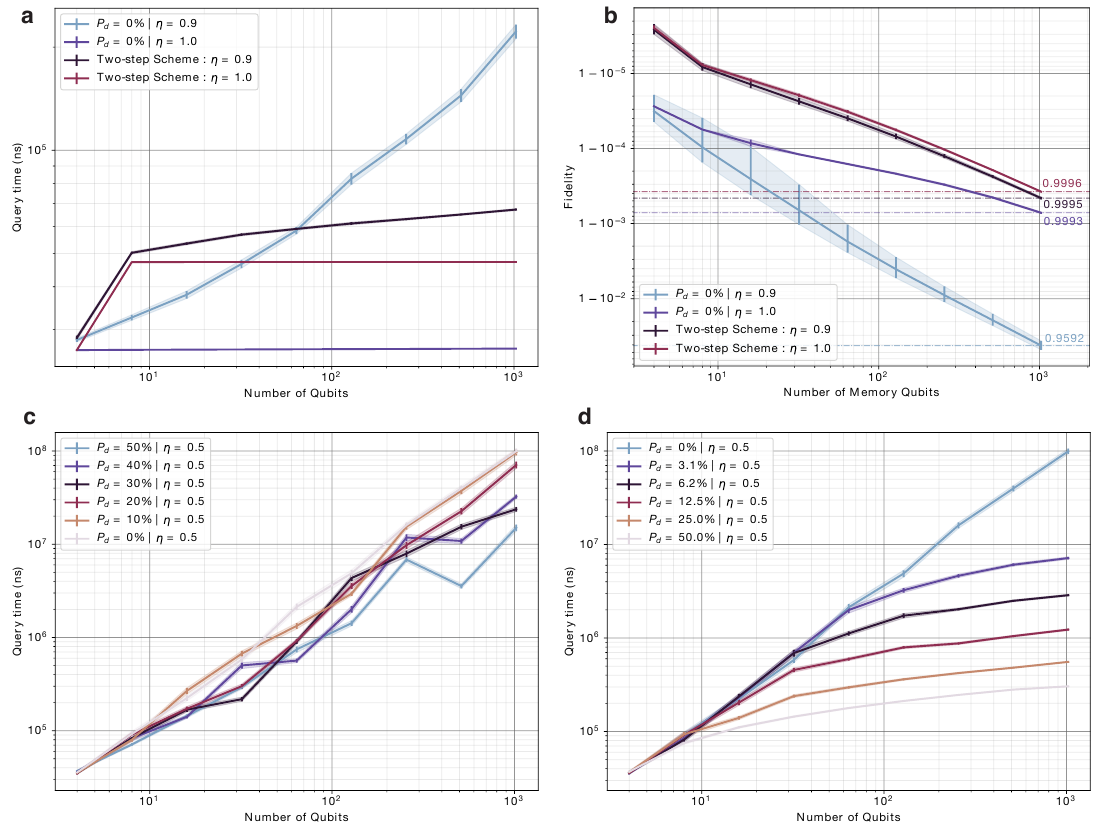}
    \caption{\textbf{TS-QRAM simulations for protocol comparison. }(a) Query times for accessing a QRAM and (b) Fidelity of access of a QRAM for a completely probabilistic hybrid scheme ($P_\mathrm{d} = 0\%$) and comparison under identical efficiencies of the distribution of Bell pairs for the TD-QRAM (two-step) scheme. $T_1 = 2~\mathrm{s}$, $T_2 = 100~\mathrm{ms}$, and $\epsilon_{CNOT} = 0$. (c) Query time scaling for randomly distributed deterministic nodes under the regular binary tree ordering and (d) and placed at higher-level steps for the linking tree, comparing between the two deterministic CNOT placement strategies for distributing the GHZ states in the TS-QRAM scheme (see Fig. \ref{fig:linkorderoptions}). Notice that for the non-random placement strategy, the ratio of deterministic nodes $P_\mathrm{d}$ is approximately given by $P_\mathrm{d} \approx 2^{-(\log_2(N)-D)-1}$. In both cases, the efficiency of the distribution of a Bell pair was set at $\eta = 0.5$. All the error bars over the data correspond to the error of the average value over 100 simulations of the protocol.}
    \label{fig:baselinecomparison} \label{fig:query_times_repeater}
\end{figure*}

\bbsection{Simulating the effects of Decoherence for a TS-QRAM}
Using the aforementioned results, we compare the TD-QRAM protocol with the TS-QRAM that includes both probabilistic EPR pair generation and deterministic linking. For comparison purposes, we start by assuming all probabilistic CNOTs ($Prob \ (\text{node being deterministic}) \equiv P_\mathrm{d} = 0\%$) have unit efficiency in the latter scheme. If the deterministic and the probabilistic CNOTs are of the same order in gate time, then the binary-tree approach is bound to be more time-consuming considering its greater number of entanglement attempts. However, the probabilistic CNOT based on cavity-reflection is typically several orders of magnitude faster than the deterministic CNOT ($10^1~\mathrm{ns}$ vs. $10^5~\mathrm{ns}$). We therefore present both the QRAM's query time and its fidelity for both schemes in Fig.~\ref{fig:baselinecomparison}, assuming perfect deterministic CNOTs. We also consider the case where the CNOT efficiency is less than unity for comparison.

Before moving onto the noise simulations, we delve into the query times. It is not obvious that now the query times scale logarithmically (or even poly-logarithmically), since the efficiency of the distributed CNOTs can increase the query times depending on the order of the linking steps. In fact, Figure~\ref{fig:baselinecomparison} already shows a non-logarithmic behavior when considering a completely probabilistic protocol. If one were to choose sequential linking steps, the query times would increase exponentially with the efficiency. By choosing the scheme demonstrated in Fig.~\ref{fig:linkorder}, we are able to reduce this to polynomial scaling~\cite{Coopmans2022}. However, depending on the noise parameters, this increase in time, compared to the initial two-step scheme, might not be wanted, as we will verify next. In Fig.~\ref{fig:query_times_repeater}, we present the query times for different efficiencies of the distributed CNOT, under the two possible hybrid schemes, with different number of deterministic CNOTs placed strategically (see Fig.~\ref{fig:linkorderoptions}).

We start by verifying that, for a random placement of the deterministic nodes, there is no clear dependence on the number of deterministic CNOTs. The reason is that, when choosing random placements for the deterministic nodes, the best order for the linking steps immediately changes and is no longer a binary tree. There already exist algorithms~\cite{Dai2020} that use linear programming to solve an identical problem of finding the best order to attempt entanglement swapping along a chain, which is virtually identical to our problem. However, the polynomial scaling of these algorithms in terms of the number of nodes of the chain makes it unsuitable for exponentially growing chains. For the intuitive placement of the deterministic nodes, this is not the case, as choosing only the top layers of the linking tree does not change the best order to do the linking. We also consider varying the efficiency of the distribution of the Bell pairs, as shown in Supplementary Figure 1.

\begin{figure}[H]
\centering
\includegraphics[width=\columnwidth]{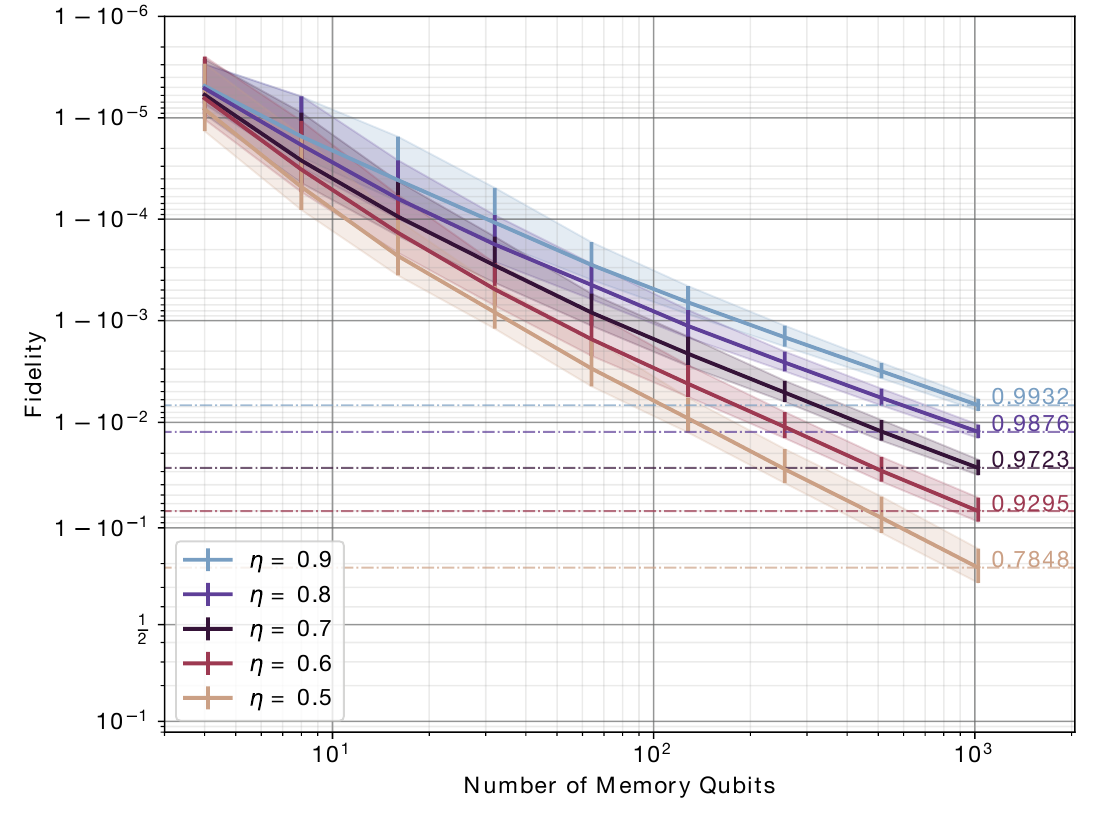}
\caption{\textbf{Fidelity scaling for a dephasing time $T_2 =1~\mathrm{s}$ and amplitude-damping time $T_1 = 2~\mathrm{s}$.} The simulations are for the completely probabilistic execution of the linking step ($P_\mathrm{d} = 0\%$), meaning there are no deterministic CNOTs being executed to create the GHZ states within each layer of the QRAM. We present different simulations for several possible values for the efficiency of each distributed CNOT ($i.e.$ the probability of success of each of the distributed CNOT), namely $\eta \in \{0.5, 0.6, 0.7, 0.8, 0.9\}$. All the error bars over the data correspond to the error of the average value over 100 simulations of the protocol.}
\label{fig:repeater_baseline}
\end{figure}

\bbsection{Dephasing and Damping Errors for a TS-QRAM}
We start by considering the case where there are no deterministic CNOTs, and vary the dephasing and damping parameters, $T_2$ and $T_1$ respectively. Note that, as expected, the query times have increased by orders of magnitude (see in greater detail in Supplementary Figure 2), hence the extent of decoherence in the memories. Moreover, to overcome the necessity of performing noisy deterministic CNOTs, the qubits used are now the electronic spin qubits, whose dephasing and damping times are much smaller than their nuclear counterparts, thereby limiting the fidelity of the QRAM tree state. For this reason, we analyzed a wide range of possible values for $T_1^\mathrm{e}$ and $T_2^\mathrm{e}$: $\{20~\mathrm{ms},\ 200~\mathrm{ms},2~\mathrm{s},20~ \mathrm{s}\}$ and $\{10~\mathrm{ms},100~\mathrm{ms},1~\mathrm{s},10~\mathrm{s}\}$, respectively. In this scenario of having only probabilistic distributed CNOTs, we analyse for multiple CNOT efficiencies $\eta$ and $T_2$ values, fixing $T_1 = 2~$s, as its contribution to the error is negligible compared to the $T_2$. In Fig.~\ref{fig:repeater_baseline}, we observe infidelity values scaling exponentially with the number of qubits for a completely probabilistic execution of the hybrid protocol. Only for memory coherence times on the second timescale, i.e. $T_2 = 1~\mathrm{s}$, does the fidelity reach around $80\%$ under a CNOT efficiency of $\eta = 0.5$. For other combinations of parameters, we refer to Supplementary Figures 3, 4 and 5.

\begin{figure*}[t]
\centering
\includegraphics[width=\textwidth]{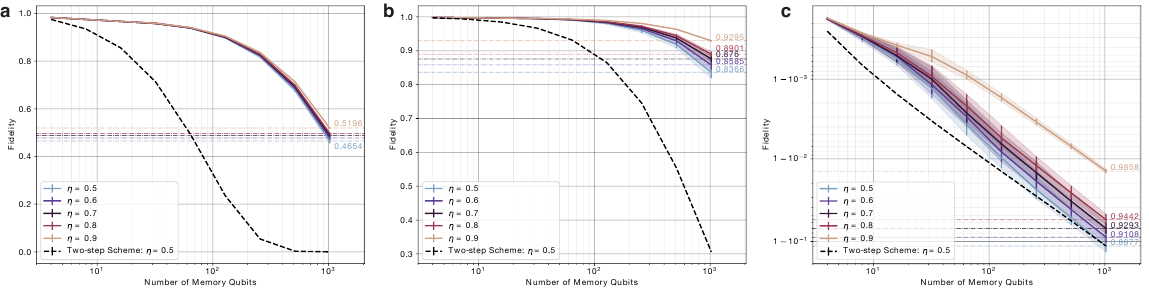}
 \caption{\textbf{Fidelity scaling for a dephasing time $T_2 = 1~\mathrm{s}$, amplitude-damping time of $T_1 = 2~\mathrm{s}$ and a varying CNOT error.} (a) $\epsilon_{\text{CNOT}} = 10^{-2}$. (b) $\epsilon_{\text{CNOT}} = 10^{-3}$. (c) $\epsilon_{\text{CNOT}} = 10^{-4}$. The simulations are in the hybrid regime, with $6.2\%$ of deterministic nodes. We present different simulations for several possible values for the efficiency of each distributed CNOT ($i.e.$ the probability of success of each of the distributed CNOT), namely $\eta \in \{0.5, 0.6, 0.7, 0.8, 0.9\}$. In dashed black lines, we also plot the values for the two-step scheme for the case with $\eta = 0.5$, and with the same $T_1,T_2$ and $\epsilon_{\text{CNOT}}$ as the hybrid scheme. All the error bars over the data correspond to the error of the average value over 100 simulations of the protocol.}
\label{fig:hybrid_fidelities}
\end{figure*}

\bbsection{Dephasing, Damping and Noisy CNOTs for TS-QRAM}
Here, we explore adding some noisy deterministic CNOTs to counteract the effect of the decoherence for longer periods of time.
As seen previously, the better location for these deterministic CNOTs are the nodes that perform the linking step at higher levels of the linking tree. In our simulations, we evaluate different values of the first deterministic layer $\log_2(N)-D \in \{2,3,4,5,6\}$. The results are presented in Fig.~\ref{fig:hybrid_fidelities}.

Depending on the CNOT error, the TS-QRAM scheme can surpass the fidelities of access of the TD-QRAM scheme under high enough $T_2$ times in the order of seconds. For other possible sets of parameters, we refer again to Supplementary Figures 3, 4 and 5. 

\bsection{Discussion}
In this article, we introduce a method to simulate large quantum networks in an open system model. Specifically, this approach enables us to model networks comprising hundreds of stationary qubits by modeling decoherence processes as noisy channels with spin dephasing errors, spin-flip errors, and noisy CNOT gates. When applied to the challenging but important problem of network-based QRAM, we find that the qubit depth of memory calls in the recently proposed TD-QRAM architecture becomes limited by CNOT errors. To overcome this bottleneck, we propose a modified network-enabled QRAM in which the noisy deterministic gates of Ref.~\cite{Nguyen2019} are replaced by \textit{heralded probabilistic} CNOT gates, which can sharply reduce gate errors. This scheme, TS-QRAM, trades increased query time for improved memory access fidelity and/or memory depth. The TS-QRAM protocol makes use of already demonstrated elements (see Supplementary Table I), suggesting the viability of near-term demonstrations in platforms of solid-state color centers as well as potentially other atomic memory modalities. 

An outstanding problem relates to the compounding loss of photonic qubits with increasing memory depth. Since teleportation-based QRAM~\cite{Chen2021} has shown that distributed quantum computers naturally map onto quantum networks, error correction schemes proposed for the former may be applied to address the issue of photon loss for the latter. Approaches include (i) photonic forward error correction using, for example, 2D photonic cluster states~\cite{Pichler2017,Larsen2019,Russo2019,Pant2019a,Uppu2020,Michaels2021} and (ii) error-corrected cluster states~\cite{Nickerson2014,Nemoto2014,Choi2019}. We leave the exploration of error correction schemes in the context of QRAM for future studies.

\bsection{Methods}
\bbsection{Discrete-time-event based simulations with NetSquid}
Given the complexity of a quantum network and its formulations, a tool such as NetSquid~\cite{Coopmans2020e} is essential to simulating a QRAM. NetSquid is capable of defining intricate discrete-time-event based protocols, with a number of steps and operations that are executed conditioned on the signaling and heralding of prior processes. Furthermore, NetSquid can simulate quantum circuits, providing methods for \textit{(i)} stabilizer circuits, with simpler and faster execution, of complexity $\order{m^2}$, where $m$ is the number of qubits; \textit{(ii)} graph states formalism, with possibly even faster execution, in $\order{d^2}$, where $\log{m} < d < m$ and $m$ is again the number of qubits; \textit{(iii)} density matrix formalism, which is slower in execution, in $\order{2^{3m}}$; \textit{(iv)} sparse density matrix formalism that relies on sparse matrix codes to speed up the execution.

Instead of using the density matrix formalism from NetSquid, we begin by retrieving  all the noise information  in each step of the protocol from a \textit{noiseless} discrete-time event simulation, resorting to the stabilizer formalism in polynomial time.  The noise information is constituted by the time qubits spent decohering, together with the information about the channels of decoherence that would have been applied in real noisy simulations, both for waiting times and gate errors. We then incorporate the extracted information to estimate the effects of decoherence at each step \textit{a posteriori}. With this information at hand, we have access to the time-evolved state of the QRAM tree at \textit{all} steps of the protocol,  which allows us to reconstruct the noise that would have been applied in the system, in a noisy simulation. To reconstruct the density matrix, we find the analytical expressions for the density matrices of smaller parts of the system, and how they evolve after the required operations, under a set of noise channels. We express these as fundamental building blocks in terms of noise parameters, namely the probability of error and time of decoherence. This is what allows to postpone the noise calculations to the end of the simulation, without losing the effects of the natural stochastic behavior of the protocol. In a way, this can be understood as pre-compiling the error effect on the intermediary states of the protocol, to shortcome the exponential complexity of calculating the density matrix at each time step using a quantum simulator.

In the rest of the methods section, we formalise these elementary building blocks for the operations required to create these GHZ states across each physical layer in the QRAM, and explain how different types of noise affect each of the intermediate steps, allowing for a reconstruction of the density matrix. We analysed dephasing, damping and depolarizing channels, and we believe other noise models could be added in a similar manner. The result is an explicit description of the final state of the QRAM access tree state prior to the execution of the teleportation protocol. The error in the state of the access tree encompasses the majority of all the error of accessing the QRAM, as the number of steps and operations made \textit{after} creating the GHZ state grows with the total number of memory addresses \textit{logarithmically}, whereas the process of generating the GHZ state requires a number of operations linear with the number of memories.

\bbsection{Elementary building blocks for TD-QRAM}
The protocol for generating GHZ states across each layer consists of two steps (see Fig.~\ref{fig:twosteps}):

\begin{figure*}[t]
\centering
\includegraphics[width=\textwidth]{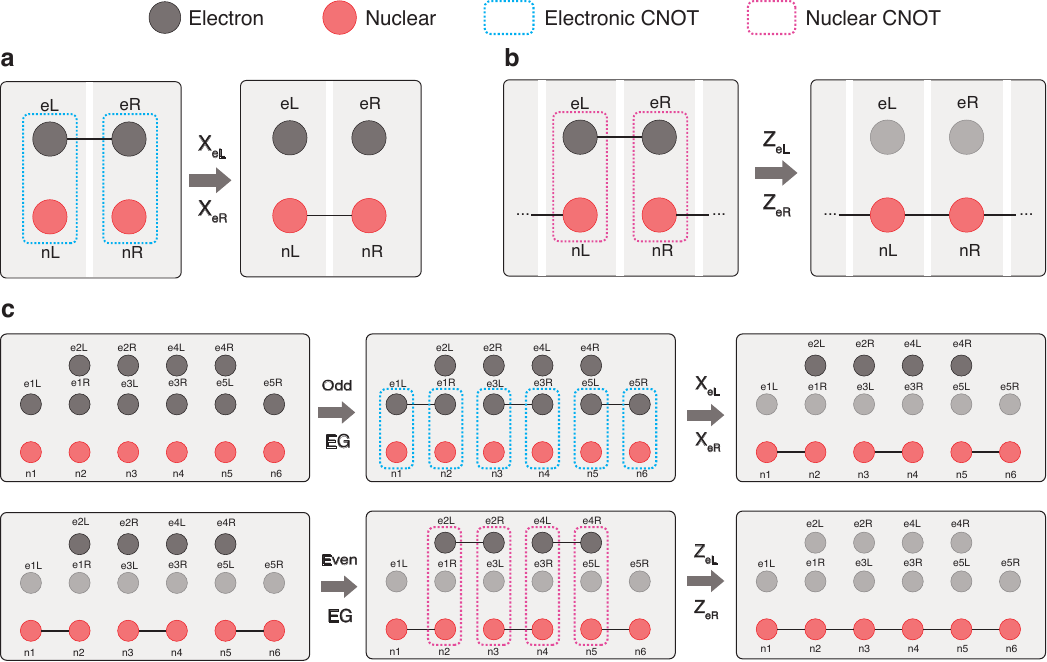}
\caption{\textbf{Building blocks for creating the GHZ state for a chain of 6-qubits} excluding the final step of entangling with the quantum computer: (a) Creating an EPR pair between two electrons, then transferring the entanglement to the nuclear spin qubits. (b) Linking two GHZ states through an entangled pair and a set of operations and measurements. (c) EPR pair creation along the odd-indexed links and transfer followed by linking of pairs by pre-sharing a EPR along the even-indexed links. Note the index of a link is with respect to the left node numbering.}
\label{fig:twosteps} \label{fig:block1} \label{fig:block2}
\end{figure*}

\begin{enumerate}
    \item Generating entanglement between the odd-indexed links. This entails first distributing photon-mediated heralded entanglement between the electrons, with a certain efficiency $\eta$, followed by \textit{electronic} CNOTs being applied with the electron qubit acting as the control and the nuclear qubits as the target. Finally, a measurement of the electron spin in the $X$ basis, with posterior corrections sent to the nuclear qubits.\\
    
    \item The second step links the entangled pairs, creating a larger GHZ state distributed across each layer. This starts off by generating heralded entanglement, with the same efficiency $\eta$, between the even-indexed  links, followed by applying \textit{nuclear} CNOTs, where now the control is the nuclear qubit and the target is the electronic qubit. We then make consequent measurements in the $Z$ basis on both electronic qubits followed by appropriate Pauli corrections.
\end{enumerate}

After the aforementioned steps, each physical layer hosts a GHZ state shared among all the nodes. Subsequently, each physical layer extends its $(2^{k-1})$-GHZ state into a $(2^{k-1}+1)$-GHZ state by sharing an additional entangled pair between an outermost node in the QRAM layer and the quantum computer, which holds the address state. After performing a Bell state measurement and corresponding corrections, the address state is teleported to the QRAM. Lastly, the memories can be accessed in superposition to complete the QRAM protocol. 

\bbsection{EPR pair creation and transferring for TD-QRAM}
The first step to creating a GHZ state across each layer is to share entanglement between neighbouring nodes. EPR pairs are created by performing a distributed CNOT gate between these nodes' electronic spin qubits, mediated by a photon. Despite the process being probabilistic with an efficiency dependent on the experimental implementation, it is a \textit{heralded} entanglement. Hence, the presence (absence) of photon detection informs the success (failure) of the entangling attempt. After this CNOT is applied between the electronic spins, an EPR pair is created and transferred to the nuclear spins in each node via a deterministic \textit{electronic} CNOT.

Given the different operations and various types of qubits involved, we introduce noise sources in the system to estimate the protocol's fidelity. We consider amplitude damping, dephasing, and CNOT gate errors for both electronic and nuclear spin qubits. 

In this step, illustrated in Fig.~\ref{fig:block1}, the following takes place:
\begin{enumerate}
    \item $\mathrm{eL}$ and $\mathrm{eR}$ decohere for a duration of time $t_\mathrm{eL}$ and $t_\mathrm{eR}$, respectively;
    \item An \textit{electronic} CNOT is applied between $\mathrm{eL}$ and $\mathrm{nL}$, with an error probability of $p_\mathrm{e}$,
    \item An \textit{electronic} CNOT is applied between $\mathrm{eR}$ and $\mathrm{nR}$ with an error probability of $p_\mathrm{e}$,
    \item $\mathrm{nL}$ and $\mathrm{nR}$ decohere for a duration of time $t_\mathrm{nL}$ and $t_\mathrm{nR}$, respectively.
\end{enumerate}

Hence, these parameters, plus the parameters associated with the physical systems, namely the $T_1$ and $T_2$ times, govern the final form of the entangled pairs. Using the notation $\epsilon(\sigma) = 1 - e^{-\sigma}$ and $\overline{\epsilon}(\sigma) = 1 - \epsilon(\sigma) = e^{-\sigma}$ for parameters that are functions of other physical parameters, namely the elapsed times and coherence times. We will use $\epsilon$ for parameters that go to zero in the absence of noise, as is the case for $\epsilon(\cdot)$ and $p_\mathrm{n}$. We also further assume $\epsilon \ll 1$. We then apply the following sequence of noise channels (check Supplementary Methods for details on the parameters): 

\begin{enumerate}
    \item Apply a Dephasing channel with probability $\epsilon(t_\mathrm{eL}/T_2^\mathrm{e}) \equiv \epsilon_\mathrm{eL}^{(2)}$ and $\epsilon(t_\mathrm{eR}/T_2^\mathrm{e}) \equiv \epsilon_\mathrm{eR}^{(2)}$ to electronic spin qubits $\mathrm{eL}$ and $\mathrm{eR}$, respectively;
    \item Apply an Amplitude damping channel with probability $\epsilon(t_\mathrm{eL}/T_1^\mathrm{e}) \equiv \epsilon_\mathrm{eL}^{(1)}$ and $\epsilon(t_\mathrm{eR}/T_1^\mathrm{e}) \equiv \epsilon_\mathrm{eR}^{(1)}$ to electronic spin qubits $\mathrm{eL}$ and $\mathrm{eR}$, respectively;
    \item Apply Depolarising channels with probability $p_{e}$ to all qubits, after applying CNOTs (modelling a noisy CNOT);
    \item Apply a Dephasing channel with probability $\epsilon(t_\mathrm{nL}/T_2^\mathrm{n}) \equiv \epsilon_\mathrm{nL}^{(2)}$ and $\epsilon(t_\mathrm{nR}/T_2^\mathrm{n}) \equiv \epsilon_\mathrm{nR}^{(2)}$ to nuclear spin qubits $\mathrm{nL}$ and $\mathrm{nR}$, respectively;
    \item Apply an Amplitude damping channel with probability $\epsilon(t_\mathrm{nL}/T_1^\mathrm{n}) \equiv \epsilon_\mathrm{nL}^{(1)}$ and $\epsilon(t_\mathrm{nR}/T_1^\mathrm{n}) \equiv \epsilon_\mathrm{nR}^{(1)}$ to nuclear spin qubits $\mathrm{nL}$ and $\mathrm{nR}$, respectively;
\end{enumerate}
The final state for each entangled pair becomes:
\begin{equation}\label{eq:finaltransfer}
\frac{1}{2}\begin{pmatrix}
1-\mu & 0 & 0 & \nu\\
0 & \mu & 0 & 0\\
0 & 0 & \mu & 0\\
\nu & 0 & 0 & 1-\mu
\end{pmatrix}
\end{equation}
where
\begin{align}
    \mu & = \frac{1-f(\epsilon_\mathrm{eL}^{(1)},\epsilon_\mathrm{eR}^{(1)})(1-p_\mathrm{e})^2 g(\epsilon_\mathrm{nL}^{(1)},\epsilon_\mathrm{nR}^{(1)}) - \epsilon_\mathrm{nL}^{(1)}\epsilon_\mathrm{nR}^{(1)}}{2},\\ 
    \nu & =  \overline{\epsilon}_\mathrm{eL}^{(2)} \cdot \overline{\epsilon}_\mathrm{eR}^{(2)}\cdot  \overline{\epsilon}_\mathrm{nL}^{(2)}\cdot \overline{\epsilon}_\mathrm{nR}^{(2)}\cdot \sqrt{
    \overline{\epsilon}_\mathrm{eL}^{(1)}\cdot \overline{\epsilon}_\mathrm{eR}^{(1)}\cdot \overline{\epsilon}_\mathrm{nL}^{(1)}\cdot \overline{\epsilon}_\mathrm{nR}^{(1)}}\cdot (1-p_\mathrm{e})^4
\end{align}
and 
\begin{equation}
\begin{aligned}
    f(\epsilon_1,\epsilon_2) &= 1-\epsilon_1 -\epsilon_2+2\epsilon_1\epsilon_2, \\
    \qquad g(\epsilon_1,\epsilon_2) &= (1-\epsilon_1)(1-\epsilon_2)
\end{aligned}
\end{equation}
For intuition regarding the $\epsilon$ function, consider the following two limits: (1) $\sigma \rightarrow 0$ in the noiseless regime where the memory coherence time goes to infinity (no decoherence) and (2) $\sigma \rightarrow \infty$ where there only exists noise and all the information is scrambled. In these limits we retrieve: $\lim_{\sigma \to 0} \epsilon(\sigma) = 0$, $\lim_{\sigma \to \infty}\epsilon(\sigma) = 1$, $\lim_{\sigma \to 0} \overline{\epsilon}(\sigma) = 1$ and $\lim_{\sigma \to \infty} \overline{\epsilon}(\sigma) = 0$. 

\bbsection{Linking of Bell pairs for TD-QRAM}
The following step is crucial to extending entanglement from bipartite to GHZ states across the entire physical layer of the QRAM. It relies on using an entangled pair to combine two GHZ states of smaller sizes into a larger GHZ state, whose number of qubits equals to the sum of each of the elementary GHZ states ($i.e.$ $n_1$-GHZ linked with a $n_2$-GHZ becomes a $(n_1+n_2)$-GHZ state).

In this step, we account for decoherence before applying CNOTs, therefore entering the previous expressions for the form of each pair. The decoherence to be analysed in this step stems from:
\begin{enumerate}
    \item A nuclear CNOT gate on $\mathrm{eL}$ and $\mathrm{nL}$ with probability of error $p_\mathrm{n}$;
    \item A nuclear CNOT gate on $\mathrm{eR}$ and $\mathrm{nR}$ with probability of error $p_\mathrm{n}$;
    \item Nuclear qubits $\mathrm{nL}$ and $\mathrm{nR}$ decohere after a CNOT for $t'$.
\end{enumerate}

Additionally, for each block, we analyse the impact of decoherence by applying the following noise channels:
\begin{enumerate}
    \item Apply depolarising channels with probability $p_{n}$ to all qubits (eL, eR, nL, nR), after applying CNOTs (modelling a noisy CNOT);
    \item Apply a dephasing channel with probability $\epsilon(t_\mathrm{nL}'/T_2^\mathrm{n}) \equiv \epsilon_\mathrm{nR}'^{(2)}$ and $\epsilon(t_\mathrm{nR}'/T_2^\mathrm{n}) \equiv \epsilon_\mathrm{nL}'^{(2)}$ to nuclear spin qubits $\mathrm{nL}$ and $\mathrm{nR}$, respectively;
    \item Apply an amplitude damping channel with probability $\epsilon(t_\mathrm{nL}'/T_1^\mathrm{n}) \equiv \epsilon_\mathrm{nR}'^{(1)}$ and $\epsilon(t_\mathrm{nR}'/T_1^\mathrm{n}) \equiv \epsilon_\mathrm{nL}'^{(1)}$ to nuclear spin qubits $\mathrm{nL}$ and $\mathrm{nR}$, respectively;
\end{enumerate}

Note that all the following calculations are now lower bounds for the fidelity, as calculation of the full analytical expressions grows exponentially with the number of qubits. Because of this, we keep only the terms up to $\order{\epsilon}$. In Supplementary Methods, we detail and test the validity of our approximations.

The final GHZ state in each layer is described by a matrix with the following form:
\begin{equation}
\frac{1}{2}\begin{pmatrix}
\rho_{00} & 0 & \dots & 0 & \rho_{01}\\
0 & \epsilon & \dots & 0 & 0\\
\vdots & 0 & \ddots & 0 & \vdots \\
0 & 0 & \dots & \epsilon & 0\\
\rho_{10} & 0 & \dots & 0 & \rho_{11}
\end{pmatrix}
\label{eq:matrixgeneral}
\end{equation}
where all $\epsilon$ terms are of at least order $\mathcal{O}(\epsilon)$ and do not contribute to infidelity, as they are orthogonal to the GHZ state. 

The diagonal elements that we consider are only the first and the last, as the remaining ones have at least $\order{\epsilon}$ and, when expanding to a larger GHZ state, contribute in $\order{\epsilon^2}$ or higher orders, hence negligibly affecting the fidelity. 

Let us first consider the form of the state after executing the linking protocol in a noiseless manner, with previously noisy states, as the ones that result from the entangling step given by Eq.~\ref{eq:finaltransfer}. Starting with the simple case of a 4-qubit GHZ state built from three states of the form of Eq.~\ref{eq:finaltransfer}, with parameters $(\mu_j,\nu_j)$, $j=1,2,3$ respectively, the final matrix is:
\begin{equation}
\frac{1}{2}\begin{pmatrix}
\overline{\mu}_1\overline{\mu}_2\overline{\mu}_3 & 0 & \dots & 0 & \nu_1 \nu_2 \nu_3\\
0 & \overline{\mu}_1\mu_2\overline{\mu}_3 & \dots & 0 & 0\\
\vdots & 0 & \ddots & 0 & \vdots \\
0 & 0 & \dots & \overline{\mu}_1\mu_2\overline{\mu}_3  & 0\\
\nu_1 \nu_2 \nu_3 & 0 & \dots & 0 & \overline{\mu}_1\overline{\mu}_2\overline{\mu}_3
\end{pmatrix}\label{eq:matrixsimple}
\end{equation}
where we, again, denote a bar over a variable as 1 minus itself, $\overline{\mu}_i \equiv 1-\mu_i$. Note that each of the $\mu_i$ comes from one of the pairs used to create the GHZ state, as these pairs are solely described by two numbers $(\mu_i,\nu_i)$ (see Eq.~\ref{eq:finaltransfer}). There exists a rule for each entry in the diagonal, which we detail in Supplementary Methods, and the same rule holds for any number of qubits of the final state. The GHZ diagonal entries then become:
\begin{equation}
\rho_{00} = \rho_{11} = \overline{\mu}_1 \overline{\mu}_2 \overline{\mu}_3 
\end{equation}

Now, adding the effect of the noisy CNOTs on the state, we calculate the diagonal terms that are shown to be identical, given by:
\begin{align}
\rho_{00}' &= \rho_{11}' = \left(1-\frac{p_\mathrm{n}}{2}\right)^2 (1-\mu_1) (1-\mu_2) (1-\mu_3) \nonumber\\
&\quad - p_\mathrm{n} \left(1-\frac{p_\mathrm{n}}{2}\right) (1-\mu_1-\frac{p_\mathrm{n}}{2}) (1-\mu_2-\frac{p_\mathrm{n}}{2}) (1-2\mu_3)\nonumber\\
&\quad + \order{\epsilon^3}\nonumber\\
&= \left[\left(1-\frac{p_\mathrm{n}}{2}\right)^2 - p_\mathrm{n} \left(1-\frac{p_\mathrm{n}}{2}\right) \right] (1-\mu_1) (1-\mu_2) (1-\mu_3)\nonumber\\
&\quad + \order{\epsilon^2} \nonumber\\
&\equiv h(p_\mathrm{n})  (1-\mu_1) (1-\mu_2) (1-\mu_3) + \order{\epsilon^2} 
\end{align}
where we recall that every term with $p_\mathrm{n}, \mu_i \ll 1$ converges to zero in the noiseless limit. For the other diagonal entries, we multiply them by $h(p_\mathrm{n})$.

Finally, incorporating memory decoherence after CNOTs, we perform another approximation. For the diagonal terms, only the damping channel plays a role. The first and last entries of the diagonal become:
\begin{equation}
\begin{aligned}
\rho_{00}'' &= h(p_\mathrm{n}) \left( \overline{\mu}_1\overline{\mu}_2\overline{\mu}_3 + \epsilon_\mathrm{nL}'^{(1)} \mu_1\mu_2\overline{\mu}_3 + \epsilon_\mathrm{nR}'^{(1)} \overline{\mu}_1\mu_2\mu_3 \right) + \order{\epsilon^4}\\
 \rho_{11}'' &= \rho_{11} (1-\epsilon_\mathrm{nL}'^{(1)})(1- \epsilon_\mathrm{nR}'^{(1)})
\end{aligned}
\end{equation}
In this approximation, the extra terms that appear for the first entry are already of order $\order{\epsilon^3}$ and could be neglected. 

Lastly, we compute the off-diagonal terms by multiplying every contribution from each noise channel applied in the correct manner. The expression is given by:
\begin{equation}
\begin{aligned}
\rho_{01}'' = \rho_{10}'' =& \nu_1 \nu_2 \nu_3 \cdot  \overline{\epsilon}_\mathrm{nL}'^{(2)} \overline{\epsilon}_\mathrm{nR}'^{(2)} \sqrt{\overline{\epsilon}_\mathrm{nL}'^{(2)} \overline{\epsilon}_\mathrm{nR}'^{(2)} } \cdot  (1-p_\mathrm{n})^2 f(p_\mathrm{n},p_\mathrm{n})
\end{aligned}
\end{equation}

When extending the linking protocol to a larger number of qubits, the expressions maintain their form. We only need to add all the terms in a similar manner as in the case of the 4-GHZ state. The complete analysis is detailed in Supplementary Methods.

Thus far, we show how three types of noise (one for the two-qubit operations and two for individual memories) influence the final state of the GHZ states generated across each physical layer of the QRAM access tree. Note that we always present the full-expressions accounting for all the noise channels. In fact, if we include a specific noise channel or a subset of what we have considered, we may simply set the parameters corresponding to other noises to zero. For example, it is straightforward to verify that setting $p_\mathrm{e}$ and $p_\mathrm{n}$ to zero and $T_1$ to infinity recovers the case for only having dephasing, thus affecting only the off-diagonal terms. The same is valid for all the other noises.

\begin{figure*}[t]
\centering
\includegraphics[width=\textwidth]{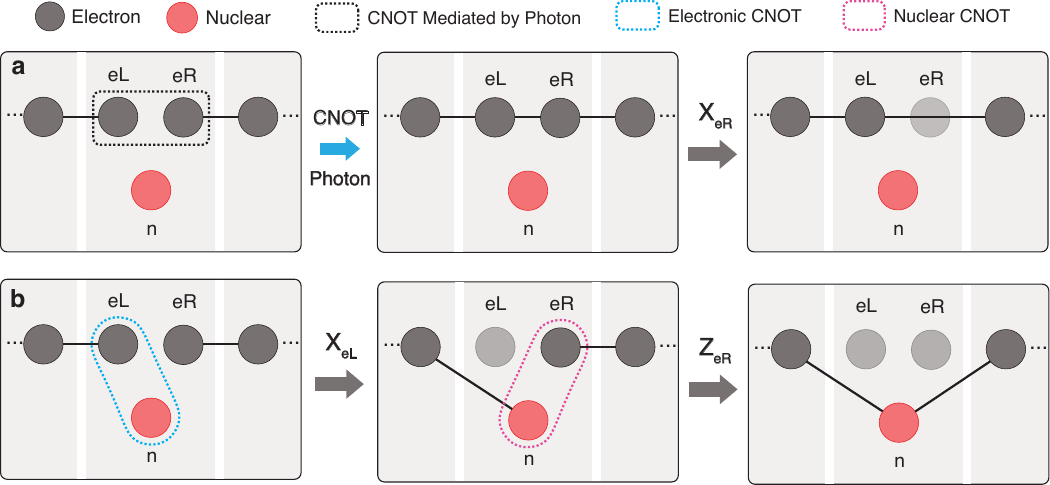}
\caption{\textbf{Possible protocols for linking smaller GHZ states into a larger GHZ state both probabilistically and deterministically.} (a) Probabilistic CNOT Protocol mediated by a photon. (b) Deterministic CNOT Protocol, consisting of a nuclear CNOT between the nuclear spin ancilla and the left electronic spin qubit.}
\label{fig:repeater_scheme}
\end{figure*}

\bbsection{Modified Protocol and Building Blocks for TS-QRAM}
We now consider an alternative architecture that enables different subsets of each layer to create GHZ states independently. As illustrated in Fig.~\ref{fig:repeater_scheme}, this architecture assumes two electron spins and one nuclear spin (instead of each node of the QRAM having an electronic spin and a nuclear spin assumed for the TD-QRAM. As we will show, this architecture still retains similar building blocks as the aforementioned TD-QRAM protocol.

\bbsection{EPR creation for TS-QRAM}
As in the non-hybrid version of the protocol, the first step to creating an EPR pair between two physically separated electronic spins is sending a photon that interacts with them sequentially. A subsequent measurement heralds the successful production of a spin-spin EPR pair. Notably, there are no deterministic CNOTs applied to transfer the qubit states onto the nuclear spins, as we only work with the electron spins at this stage.

The final state shared between the electronic spins is the one of Eq. \ref{eq:finaltransfer} in the limit of the absence of \textit{electronic} CNOT error ($p_\mathrm{e} \rightarrow 0 $) and altering the memory decoherence noise from nuclear to electronic ($\epsilon_\mathrm{nL} \rightarrow \epsilon_\mathrm{eL}'$ and $\epsilon_\mathrm{nR} \rightarrow \epsilon_\mathrm{eR}'$):
\begin{equation}
\frac{1}{2}
\begin{pmatrix}
1-\mu & 0 & 0 & \nu\\
0 & \mu & 0 & 0\\
0 & 0 & \mu & 0\\
\nu & 0 & 0 & 1-\mu
\end{pmatrix}
\label{eq:eprdensity2}
\end{equation}
where
\begin{align}
    \mu & = \frac{1-f(\epsilon_\mathrm{eL}^{(1)},\epsilon_\mathrm{eR}^{(1)}) g(\epsilon_\mathrm{eL}'^{(1)},\epsilon_\mathrm{eR}'^{(1)}) + \epsilon_\mathrm{eL}'^{(1)}\epsilon_\mathrm{eR}'^{(1)}}{2},\\ 
    \nu & =  \overline{\epsilon}_\mathrm{eL}^{(2)} \cdot \overline{\epsilon}_\mathrm{eR}^{(2)}\cdot  \overline{\epsilon}_\mathrm{eL}'^{(2)}\cdot \overline{\epsilon}_\mathrm{eR}'^{(2)}\cdot \sqrt{
    \overline{\epsilon}_\mathrm{eL}^{(1)}\cdot \overline{\epsilon}_\mathrm{eR}^{(1)}\cdot \overline{\epsilon}_\mathrm{eL}'^{(1)}\cdot \overline{\epsilon}_\mathrm{eR}'^{(1)}}
\end{align}
and again,
\begin{equation}
\begin{aligned}
    f(\epsilon_1,\epsilon_2) &= 1-\epsilon_1 -\epsilon_2+2\epsilon_1\epsilon_2, \\
    \qquad g(\epsilon_1,\epsilon_2) &= (1-\epsilon_1)(1-\epsilon_2)
\end{aligned},
\end{equation}
where we use the same abbreviation $\overline{\epsilon}(\sigma) = 1 - \epsilon(\sigma) = e^{-\sigma} $.

\bbsection{Linking Pairs in the Probabilistic Scenario for a TS-QRAM}
TS-QRAM differs from TD-QRAM in that the operation of linking pairs has a non-unity probability of succeeding - let us call this probability $p_{CNOT}$. Moreover, it is executed in a similar way as that of creating an EPR pair:
\begin{enumerate}
    \item Interact photon $\gamma$ with the left electronic spin qubit $\mathrm{eL}$, executing a local CNOT,
    \item Send the single photon $\gamma$ to the right cavity,
    \item Interact the photon $\gamma$ with the right electronic spin qubit $\mathrm{eR}$, executing a local CNOT,
    \item Measure the photon $\gamma$,
    \item Measure the right (or left) electronic spin in $X$.
\end{enumerate}

Importantly, both cavities \textit{belong to the same node} in this step. This results in a controlled gate applied between the right and left electronic spin qubits. Unlike before, it is still necessary to measure one of the nodes' electronic spin qubits, as the state has twice the number of qubits as the final state (we chose to measure the right electron, but one could choose to keep the right and measure the left instead; the choice is arbitrary and translates to the same practical outcome). This measurement should be in the $X$ basis, in order to not destroy the entanglement shared among all the qubits and rendering the state useless. Moreover, a correction must be made depending on the outcome on the measurement of the electronic spin qubit and the photonic qubit.

Afterwards, the GHZ states shared between the left and right nodes are linked into a larger GHZ state, via an intermediary node. Inside this intermediary node, its left electronic spin merges into the larger GHZ state. We again take into account the previous calculations for detailing the density matrix of the final state. The decoherence steps are now only provenient from the memories of where each qubit is being held (which we chose to be the left cavity of the node). As we used near-perfect probabilistic CNOTs mediated by a photon, only its memory affects the state fidelity. Thus, for the remainder of the protocol, we:
\begin{enumerate}
    \item Apply a dephasing channel with probability $\epsilon(t_\mathrm{eL}''/T_2^\mathrm{e}) \equiv \epsilon_\mathrm{eL}''^{(2)}$ to electronic spin qubit $\mathrm{eL}$;
    \item Apply an amplitude damping channel with probability $\epsilon(t_\mathrm{eL}''/T_1^\mathrm{e}) \equiv \epsilon_\mathrm{eL}''^{(1)}$ to electronic spin qubit $\mathrm{eL}$.
\end{enumerate}

Importantly, the following calculations are again lower-bound approximations for the fidelity. Performing the calculations for a simple link of two entangled pairs described by Eq.~\ref{eq:eprdensity2}, with parameters $(\mu_j,\nu_j), j=1,2$, the final density matrix of the 3-GHZ state, prior to any memory decoherence, is:
\begin{equation}
\frac{1}{2}\begin{pmatrix}
\overline{\mu}_1\overline{\mu}_2 & 0 & \dots & 0 & \nu_1 \nu_2 \\
0 & \overline{\mu}_1\mu_2 & \dots & 0 & 0\\
\vdots & 0 & \ddots & 0 & \vdots \\
0 & 0 & \dots & \overline{\mu}_1\mu_2  & 0\\
\nu_1 \nu_2  & 0 & \dots & 0 & \overline{\mu}_1\overline{\mu}_2
\end{pmatrix}\label{eq:matrix1prob}
\end{equation}

Adding the memory decoherence accounting for both dephasing and amplitude-damping leads to a matrix similar in form to one shown in Eq.~\ref{eq:matrixgeneral}, except with entries changing to:
\begin{equation}
\begin{aligned}
\rho_{00}' &= \overline{\mu}_1\overline{\mu}_2+ \epsilon_\mathrm{eL}''^{(1)} \mu_1\mu_2 + \order{\epsilon^4}\\
 \rho_{11}' &= \overline{\mu}_1\overline{\mu}_2 (1-\epsilon_\mathrm{eL}''^{(1)}) \\
\rho_{10}' &= \rho_{01}' = \nu_1 \nu_2 \overline{\epsilon}_\mathrm{eL}''^{(2)} \sqrt{1-\epsilon_\mathrm{eL}''^{(1)}}
\end{aligned}
\end{equation}

\bbsection{Linking Pairs in the Deterministic Scenario for TS-QRAM}
We next describe a scheme for deterministic CNOT gates. Since the composition of each node is now different, the operations one needs to execute to deterministically link smaller GHZ states into larger GHZ states changes as well:

\begin{enumerate}
    \item Apply a deterministic \textit{electronic} CNOT controlled by the left electronic spin qubit $\mathrm{eL}$ and targeted at the nuclear spin qubit $n$,
    \item Measure the left electronic spin qubit $\mathrm{eL}$ in $X$,
    \item Apply a deterministic nuclear CNOT controlled by the nuclear spin qubit $n$ and targeted at the right electronic spin qubit $\mathrm{eR}$,
    \item Measure the right electronic spin qubit $\mathrm{eR}$ in $Z$.
\end{enumerate}

Measurement-conditioned corrections result in a GHZ state consisting the nuclear spin and the remaining electronic spin qubits. Notice that by not involving the photon mediated CNOT, this has been done in a deterministic fashion. In this case, we must consider additional errors, namely those that arise from using deterministic \textit{electronic} and \textit{nuclear} CNOTs. The sequence of noise channels becomes:

\begin{enumerate}
    \item Apply depolarising channels with probability $p_\mathrm{e}$ to the electronic spin qubit $\mathrm{eL}$ and to the nuclear spin qubit $n$,
    \item Apply depolarising channels with probability $p_\mathrm{n}$ to the nuclear spin qubit $n$ and to the electronic spin qubit $\mathrm{eR}$,
    \item Apply a dephasing channel with probability $\epsilon(t_{n}'/T_2^\mathrm{n}) \equiv \epsilon_{n}'^{(2)}$ to nuclear spin qubit $n$,
    \item Apply an amplitude damping channel with probability $\epsilon(t_{n}'/T_1^\mathrm{n}) \equiv \epsilon_{n}'^{(1)}$ to nuclear spin qubit $n$.
\end{enumerate}

From here, we calculate the final state's density matrix. Performing the calculations for the same simple link of two entangled pairs described by Eq.~\ref{eq:eprdensity2}, with parameters $(\mu_j,\nu_j), j=1,2$, the final density matrix of the 3-GHZ state, prior to any memory decoherence and without any CNOT errors is the same as Eq.~\ref{eq:matrix1prob}. Adding the effect of the CNOTs leads to:

\begin{equation}
\begin{aligned}
\rho_{00}' &= \rho_{11}' =  (1-p)^2\overline{\mu}_1\overline{\mu}_2+ \frac{p}{2}\left(1-\frac{p}{2} \right)\\
\rho_{10}' &= \rho_{01}' = \nu_1 \nu_2 (1-p)^3 \left(1-\frac{p}{2} \right)
\end{aligned}
\label{eq:repeaterdeterministiccnots}
\end{equation}
where we set $p_\mathrm{e} = p_\mathrm{n} \equiv p$. In fact, all diagonal entries can be decomposed into terms of the form $(1-p)^2 \text{diag}(\rho) + \mathbb{1} p/2(1-p/2)$. Using this fact, we incorporate the posterior amplitude-damping noise channels:
\begin{equation}
\begin{aligned}
\rho_{00}'' &= \tilde{h} (p) \cdot (\overline{\mu}_1\overline{\mu}_2+ \epsilon_\mathrm{eL}'^{(1)} \mu_1\mu_2)  + \order{\epsilon^3} \\
 \rho_{11}'' &= \tilde{h} (p) \cdot \overline{\mu}_1\overline{\mu}_2 (1-\epsilon_\mathrm{eL}'^{(1)}) + \order{\epsilon^3} \\
\rho_{10}'' &= \rho_{01}'' = \nu_1 \nu_2 (1-p)^3 \left(1-\frac{p}{2} \right) \sqrt{1-\epsilon_\mathrm{eL}'^{(1)}}
\end{aligned}
\end{equation}
where $\tilde{h}(p) = (1-p)^2 + p/2 (1-p/2)$. In the Supplementary Methods we present the derivation for a chain of an arbitrary number of channels, as well as a proof of the validity of our approximations.

\begin{figure}[t]
    \centering
    \includegraphics[width=\columnwidth]{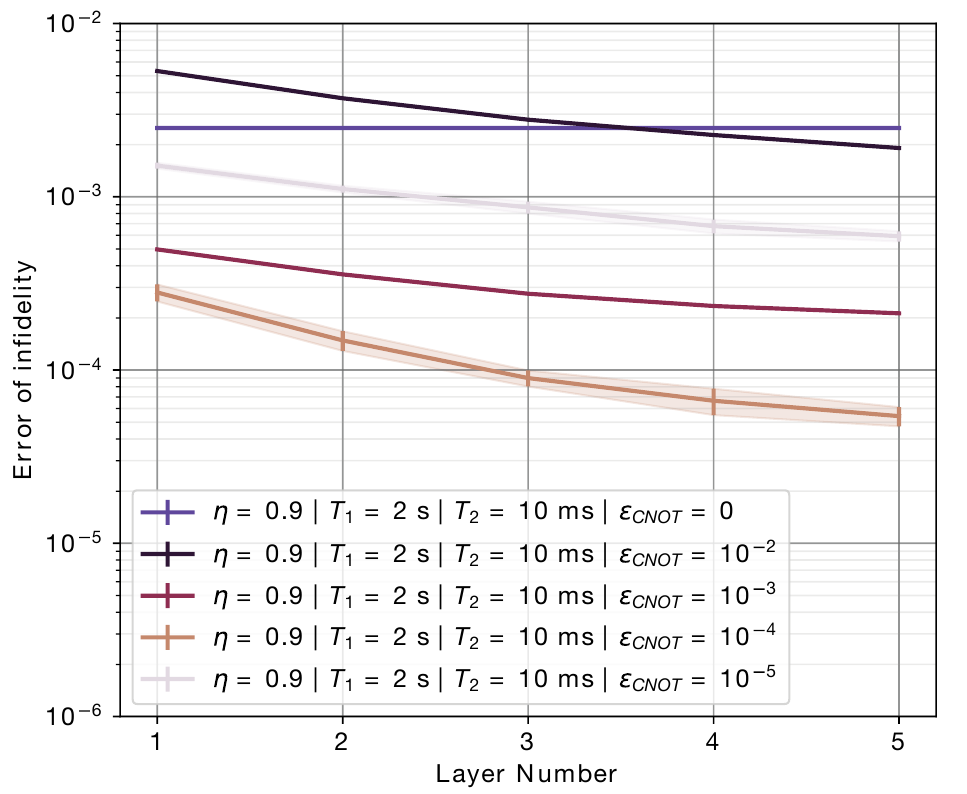}
    \caption{\textbf{Comparison between the infidelity calculated using a full density matrix simulation and our methods}. Simulations made under a TD-QRAM protocol execution up to 5 layers, for different combinations of error parameters. All the error bars over the data correspond to the error of the average value over 100 simulations of the protocol.}
    \label{fig:test}
\end{figure}

\bbsection{Validity of the Results}
In this section, we discuss our methods and provide empirical proof for our noise analysis robustness. In our work we have considered an analysis of a discrete protocol, where each gate takes a fixed amount of time and qubits stay in memory waiting for instructions from the protocol. Given the discreteness of the problem, solving the master-equations in each of the specific time periods where the noise actually happens, is equivalent to applying the corresponding noise channels. 

To support our claims and demonstrate the equivalence of using the full density matrix and our methods of postponing the noise analysis to the end of the protocol simulation, we present one additional figure. This figure shows the comparison of the fidelity of the GHZ states distributed at each of the layers of the QRAM for a small QRAM ($5$ layers, $\sim 36$ qubits) obtained from our simulation methods and those derived using the density matrix formalism. We present the comparison for simulations for the TD-QRAM, as for TS-QRAM we use the same methods, just under different assumptions over which noise channels are applied. 
Let $F_{\mathrm{DM}}$ be the fidelity calculated using the full density matrix and $F_{\mathrm{S}}$ be the one from our simulation methods. The figure showcases how the difference of the fidelities calculated using the density matrix and our methods evolve for different layers (with increasing amounts of qubits), weighted by the overall error (or infidelity of the state), \textit{i.e.} $|F_{\mathrm{DM}} - F_{\mathrm{S}}|/(1-F_{\mathrm{DM}})$.

One can observe in Fig.\ref{fig:test} that the error is always less than $1\%$, showing a trend of either maintaining or decreasing as the size of the system increases. This provides empirical proof of the power of our methods across the different scenarios.

\bsection{Data Availability}
The data supporting the results in this work is available from L.B. at \href{https://github.com/luisbugalho/HeraldedQRAM}{https://github.com/luisbugalho/HeraldedQRAM}.

\bsection{Code Availability}
The code supporting the results in this work are available from L.B. at \href{https://github.com/luisbugalho/HeraldedQRAM}{https://github.com/luisbugalho/HeraldedQRAM}.

\bsection{Acknowledgments}
L.B., E.Z.C., and Y.O. thank the support from Funda\c c\~ao para a Ci\^encia e a Tecnologia (FCT, Portugal), namely through projects UIDB/04540/2020 and UIDB/50008/2020. L.B. acknowledges the support of FCT through scholarship BD/05268/2021 and of the PEPR integrated project EPiQ ANR-22-PETQ-0007 part of Plan France 2030. W.D. is supported by the National Science Foundation to the Computing Research Association for the CIFellows 2020 Program. K.C.C. and D.E. acknowledge funding support by the National Science Foundation (NSF) Engineering Research Center for Quantum Networks (CQN), awarded under cooperative agreement number 1941583, and the ARO MURI on `Theory and Engineering of Large-Scale Distributed Entanglement'(W911NF2110325). D.E. further acknowledges support from the NSF C-Accel program, grant number 2040695.

\bsection{Author Contributions}
L.B., E.Z.C., W.D. and K.C. performed the simulation of the protocol and consequent analysis. All authors discussed the analysis of the data, and contributed to writing or proofreading the manuscript. Y.O. and E.D. supervised the project.

\bsection{Competing Interests}
The authors declare no competing financial or non-financial interests.

\bsection{Additional information}
Supplementary Information is available for this paper.

\clearpage

\onecolumngrid
\appendix

\renewcommand{\figurename}{Supplementary Figure}
\renewcommand{\tablename}{Supplementary Table}
\renewcommand\theequation{\arabic{equation}}
\renewcommand{\bsection}[1]{\noindent\vspace{0.5cm} \begin{Large}\textbf{#1} \end{Large} \par}
\renewcommand{\bbsection}[1]{\noindent\vspace{0.3cm}\begin{large}\textbf{#1.} \end{large} }
\renewcommand{\bbbsection}[1]{ \par \vspace{0.3cm} \noindent \begin{large}\textit{#1.} \end{large} }

\bsection{Supplementary Methods}

\bbsection{Brief description of noise channels}

A noise channel is a special type of quantum channel, and, as all quantum channels, it can be described by a set of Kraus operators $\{K_i\}_{\mathcal{E}}$. Its action on a quantum state, represented by a density matrix $\rho$, is given by the following equation, which can also be conceived as a unitary evolution of the system in question $\rho$ and a complementary system $\rho_E$, usually the environment, after tracing out the complementary system:
\begin{equation}
    \mathcal{E}(\rho) = \sum_i K_i^\dagger \rho K_i  \qquad \Longleftrightarrow \qquad \mathcal{E} (\rho) = \text{Tr}_E \left[ U \rho \otimes \rho_E U^\dagger \right].
\end{equation}
Because of this, these operations are completely positive trace preserving (CPTP) maps and verify the identity $\sum_i \hat{K}_i^\dagger \hat{K}_i = \mathbb{I}$. Moreover, the set of Kraus operators is not unique to a quantum channel, as a unitary map between two sets of Kraus operators leaves the action of the quantum channel itself invariant. 

As we described in the main text, we deal with several types of noise, all of them described by corresponding noise channels. In the memories we consider dephasing and amplitude-damping noise, described by corresponding dephasing and amplitude damping channels, with sets of Kraus operators given respectively by:

\begin{align}
    \mathcal{E}^{\mathrm{Deph}} &\sim \left\{ \sqrt{1-p} \ \mathbb{I}, \sqrt{p} \ \hat{Z} \right\} = \left\{ 
    \begin{bmatrix}
    \sqrt{1-p} & 0 \\
    0 & \sqrt{1-p}
    \end{bmatrix},
    \begin{bmatrix}
    \sqrt{p} & 0 \\
    0 & - \sqrt{p}
    \end{bmatrix} \right\} \\
    \mathcal{E}^{\mathrm{Damp}} &\sim \left\{|0\rangle\langle 0|+\sqrt{1-p}| 1\rangle\langle 1|, \sqrt{p}| 0\rangle\langle 1|\right\} = \left\{ 
    \begin{bmatrix}
    1 & 0 \\
    0 & \sqrt{1-p}
    \end{bmatrix},
    \begin{bmatrix}
    0 & \sqrt{p} \\
    0 & 0
    \end{bmatrix}
    \right\}
\end{align}

Moreover, to also consider the possibility of noisy CNOTs, and following the lines of NetSquid to be able to corroborate  results, we model the error in CNOTs by applying a perfect CNOT gate proceeded by two depolarising channels, one applied to the control qubit, and another applied to the target qubit:
\begin{equation}
    \tilde{\text{CNOT}} = \mathcal{E}^{\mathrm{Dep}}_i \circ \mathcal{E}^{\mathrm{Dep}}_j \circ \mathrm{CNOT}_{ij} (\rho),
\end{equation}
where the depolarising channel is defined by the following set of operators:
\begin{equation}
\begin{aligned}
    \mathcal{E}^{\mathrm{Dep}} \sim \left\{ \sqrt{1-p} \ \mathbb{I}, \sqrt{p/3} \ \hat{X}, \sqrt{p/3} \ \hat{Y}, \sqrt{p/3} \ \hat{Z} \right\} = \Bigg\{ &
    \begin{bmatrix}
    \sqrt{1-p} & 0 \\
    0 & \sqrt{1-p}
    \end{bmatrix},
    \begin{bmatrix}
    0 & \sqrt{p/3}  \\
    \sqrt{p/3} & 0
    \end{bmatrix}, \\
    &\begin{bmatrix}
    0 & - i \sqrt{p/3} \\
    i \sqrt{p/3} & 0
    \end{bmatrix},
    \begin{bmatrix}
    \sqrt{p} & 0 \\
    0 & - \sqrt{p}
    \end{bmatrix} \Bigg\} 
\end{aligned}
\end{equation}

\bbsection{Bell Pair States under different Noise Channels}

For completion let us describe the memory noise channels action on the elementary building block of either of the detailed schemes, namely, the Bell pair state  $\ket{\phi^+} \propto \ket{00} + \ket{11}$. We do this for an arbitrary error parameter $p$, and one that is useful for the noise analysis in terms of the time the qubits spend decohering in the memories $\sigma = \Delta t / T_{coh}$.

\begin{enumerate}
	\item Dephasing:
	\begin{equation}
	\begin{aligned}
		\mathcal{E}_1^{\text{Deph}} (\ket{\phi^+}\bra{\phi^+}) &= \mathcal{E}_2^{\text{Deph}} (\ket{\phi^+}\bra{\phi^+}) \\
		&= (1-p) \ket{\phi^+}\bra{\phi^+} + p \ket{\phi^-}\bra{\phi^-} \\
		&=\frac{1}{2}
        \begin{bmatrix}
         1 & 0 & 0 & 1-2p \\
         0 & 0 & 0 & 0 \\
         0 & 0 & 0 & 0 \\
         1-2p & 0 & 0 & 1 \\
        \end{bmatrix} \quad \overset{2p \rightarrow 1-e^{-\sigma}}{\xrightarrow{\hspace*{1.5cm}}} \quad 
        \frac{1}{2}
        \begin{bmatrix}
         1 & 0 & 0 & e^{-\sigma} \\
         0 & 0 & 0 & 0 \\
         0 & 0 & 0 & 0 \\
         e^{-\sigma} & 0 & 0 & 1 \\
        \end{bmatrix}
	\end{aligned}
	\end{equation}
	\item Damping: 
	\begin{equation}
	\begin{aligned}
		\mathcal{E}_1^{\text{Damp}} (\ket{\phi^+}\bra{\phi^+}) &=\frac{1}{2}
        \begin{bmatrix}
         1 & 0 & 0 & \sqrt{1-p} \\
         0 & 0 & 0 & 0 \\
         0 & 0 & p & 0 \\
         \sqrt{1-p} & 0 & 0 & 1-p \\
        \end{bmatrix} 
        \quad \overset{p \rightarrow 1-e^{-\sigma}}{\xrightarrow{\hspace*{1.5cm}}} \quad 
        \frac{1}{2}
        \begin{bmatrix}
         1 & 0 & 0 & e^{-\sigma/2} \\
         0 & 0 & 0 & 0 \\
         0 & 0 & 1-e^{-\sigma} & 0 \\
        e^{-\sigma/2} & 0 & 0 & e^{-\sigma} \\
        \end{bmatrix} 
        \\
        \mathcal{E}_2^{\text{Damp}} (\ket{\phi^+}\bra{\phi^+}) &=\frac{1}{2}
        \begin{bmatrix}
         1 & 0 & 0 & \sqrt{1-p} \\
         0 & p & 0 & 0 \\
         0 & 0 & 0 & 0 \\
         \sqrt{1-p} & 0 & 0 & 1-p \\
        \end{bmatrix} 
        \quad \overset{p \rightarrow 1-e^{-\sigma}}{\xrightarrow{\hspace*{1.5cm}}} \quad 
        \frac{1}{2}
        \begin{bmatrix}
         1 & 0 & 0 & e^{-\sigma/2} \\
         0 & 1-e^{-\sigma} & 0 & 0 \\
         0 & 0 & 0 & 0 \\
        e^{-\sigma/2} & 0 & 0 & e^{-\sigma} \\
        \end{bmatrix} 
	\end{aligned}
	\end{equation}
\end{enumerate}

\bbsection{Detailed Calculations and Validity of Approximations \label{appendix:approximations}}

In this appendix we detail all the calculations for extending the previous results for arbitrary number of qubits GHZ states, together with verifying the approximations made throughout the calculations, namely for the amplitude-damping channel and the noisy CNOTs. To do this we utilise some results known from quantum channels theory, that simplify the result into something easier to handle. 

The first important result is the following: let $\mathcal{E} (\cdot)$ and $\mathcal{D} (\cdot)$ be two arbitrary noise channels, with corresponding sets of Kraus operators $\{\hat{K}_i\}$ and $\{\hat{M}_i\}$:
\begin{equation}
    \mathcal{E} \circ \mathcal{D} \ (\cdot) = \mathcal{D} \circ \mathcal{E} \ (\cdot)
\end{equation}
if $[\hat{K}_i,\hat{M}_j]=0 \  \forall i,j$, meaning, all Kraus operators commute. The most obvious case of this is the case where $\mathcal{E}(\cdot)$ acts over a subsystem $A$ and $\mathcal{D}(\cdot)$ acts over a subsystem $B$, such that $A \cap B = \emptyset$, which is always the case when applying noise channels over distinct qubits.

This will allow in particular to refrain from calculating the final memory decoherence until the final step, and divide both protocols always in three phases: Bell pair creation, linking the GHZ states (whether they are deterministic as in the teleportation-based protocol, or probabilistic, in the hybrid protocol), and final decoherence over the memories.

\bbbsection{Bell Pair Creation}
The first step of Bell pair creation is already detailed in the main text. To verify this is simply to apply every noise channel described in the text following the protocol, and symmetrize the state, since the amplitude-damping channel introduces some terms which are not symmetric. Later in the Supplementary Methods, we explain exactly what we mean by symmetrizing, and the correspondence within the protocol.

\bbbsection{Linking GHZ States}
For the linking part, we assume that every pair that will be linked is described by:
\begin{equation*}
\frac{1}{2}\begin{pmatrix}
1-\mu_i & 0 & 0 & \nu_i\\
0 & \mu_i & 0 & 0\\
0 & 0 & \mu_i & 0\\
\nu_i & 0 & 0 & 1-\mu_i
\end{pmatrix},
\end{equation*}
where the $\mu_i$s and the $\nu_i$s depend on all the errors previous to the entanglement linking step. 
Let us start by the most simple case: the one where the linking does not contain any error, meaning no noisy deterministic CNOTs are applied. Take two different pairs on qubits $i_1,i_2$ and $j_1,j_2$:
\begin{equation*}
\frac{1}{2}\begin{pmatrix}
1-\mu_1 & 0 & 0 & \nu_1\\
0 & \mu_1 & 0 & 0\\
0 & 0 & \mu_1 & 0\\
\nu_1 & 0 & 0 & 1-\mu_1
\end{pmatrix} \qquad , \qquad
\frac{1}{2}\begin{pmatrix}
1-\mu_2 & 0 & 0 & \nu_2\\
0 & \mu_2 & 0 & 0\\
0 & 0 & \mu_2 & 0\\
\nu_2 & 0 & 0 & 1-\mu_2
\end{pmatrix}
\end{equation*}
The protocol, under this assumption, is identical to applying a regular CNOT between qubits $i_2$ and $j_1$ and measuring afterwards either qubit $i_2$ or $j_1$ in the $X$ basis, with posterior correction. The result of doing so is:
\begin{equation}
 \frac{1}{2} \left(
\begin{array}{cccccccc}
 \left(1-\mu_1\right) \left(1-\mu_2\right) & 0 & 0 & 0 & 0 & 0 & 0 & \nu _1 \nu _2 \\
 0 & \left(1-\mu_1\right) \mu_2 & 0 & 0 & 0 & 0 & 0 & 0 \\
 0 & 0 & \mu _1 \mu _2 & 0 & 0 & 0 & 0 & 0 \\
 0 & 0 & 0 & \mu _1 \left(1-\mu _2\right) & 0 & 0 & 0 & 0 \\
 0 & 0 & 0 & 0 &  \mu _1 \left(1-\mu _2\right) & 0 & 0 & 0 \\
 0 & 0 & 0 & 0 & 0 & \mu _1 \mu _2 & 0 & 0 \\
 0 & 0 & 0 & 0 & 0 & 0 & \left(1-\mu _1\right) \mu _2 & 0 \\
 \nu _1 \nu _2 & 0 & 0 & 0 & 0 & 0 & 0 &  \left(1-\mu _1\right) \left(1-\mu _2\right) \\
\end{array}
\right)
\end{equation}
In fact, we can see the general rule from this already simple example: the entry $\rho_{ii}$ of the density matrix of the GHZ state connecting $n$ nodes is given by:
\begin{equation}
\begin{aligned}
    \rho_{ii} &= \prod_{(j,j+1) \in L(n)} \theta_i ( \mu_j) \ ,  \text{ where }
    \theta_i (\mu_j) &= \begin{cases}
    1-\mu_j &, \text{if } i_j + i_{j+1} \text{ mod } 2 = 0 \\
    \mu_j &, \text{otherwise}
    \end{cases} \\
    \rho_{01} &= \rho_{10} = \prod_{(j,j+1) \in L(n)} \nu_j
\end{aligned}
\label{eq:cnotslinkingformula}
\end{equation}
where $L(n)$ represents the line graph composed by the set of vertices $(i,i+1)$, $1 \leq i \leq n-1$, and where $(i,j)$ represents one link, equivalently one Bell pair, between one qubit on node $i$ and one qubit on node $j$. Moreover, consider the notation $i_k$ as the $k$th most valuable bit of the number $i$ expressed in basis 2, with word length $n$. For example $5_3$ with word length 4 is given by first converting 5 to basis 2, $5 = 0101$, and then choosing its 3rd most valuable digit counting from the right: $5_3 = 1$. This is not really that surprising, as applying CNOTs can be seen as an addition modulo 2 of two qubits expressed in the computational basis. Since the CNOTs are applied between neighbouring links, Eq. \ref{eq:cnotslinkingformula} seems somewhat intuitive.

\bbbsection{Noisy CNOTs -- Two-Step Protocol}
Finally, the only step missing here is the step of adding the noise contribution of the CNOTs. To do this, we have to consider separately the cases for the two-step protocol and the repeater protocol (only the deterministic execution, as it is the one where noisy CNOTs are applied). The simpler way to calculate this is recursively. 
For the two-step protocol we had that:

\begin{equation*}
\frac{1}{2}\begin{pmatrix}
\rho_{00} & 0 & \dots & 0 & \rho_{01}\\
0 & \epsilon & \dots & 0 & 0\\
\vdots & 0 & \ddots & 0 & \vdots \\
0 & 0 & \dots & \epsilon & 0\\
\rho_{10} & 0 & \dots & 0 & \rho_{11}
\end{pmatrix},
\end{equation*}
where $\rho_{00} = \rho_{11} = h(p_n)  (1-\mu_1) (1-\mu_2) (1-\mu_3) + \order{\epsilon^2} $ and in fact the diagonal terms that are not part of the GHZ state are always at least $\order{\epsilon^1}$. The first thing to note is that, under the linking protocol, these diagonal terms can never jump to the GHZ state entries without it resulting from an error, meaning, they always contribute to the fidelity of the final state in $\order{\epsilon^1}\cdot \order{\epsilon^1} = \order{\epsilon^2}$. Doing so and calculating the GHZ entries recursively, by adding more links, we get that:
\begin{equation}
\begin{aligned}
    \rho_{00}' = \rho_{11}' &= \rho_{00} \cdot \prod_{(j,j+1) \in L^{\mathrm{E}}(n)} h(p) + \order{\epsilon^2} \\
    &= \rho_{00} \cdot \left[h(p)\right]^{n/2} + \order{\epsilon^2} \\
    h(p_n) &= \left(1-\frac{p}{2}\right)^2 - p \left(1-\frac{p}{2}\right)
\end{aligned},
\label{eq:noisycnotslinkingtwostep}
\end{equation}
where $p \equiv p_n$ it is the CNOT error, and we denote by $L^{\mathrm{E}}(n)$ the set of links in $L(n)$ with links $(j,j+1)$ where $j$ is even. Since in a chain with $n$ links, half of them are even, assuming all the CNOTs have an equal amount of noise, the first expression can be simplified into the second expression of Eq. \ref{eq:noisycnotslinkingtwostep}.

As for the off-diagonal terms, namely the corner entries corresponding to the GHZ state, as usual, the expression is quite simple and an exact value. They will be given by:

\begin{equation}
\begin{aligned}
    \rho_{01}' = \rho_{10}' &= \rho_{01} \cdot \prod_{(j,j+1) \in L^{\mathrm{E}}(n)} (1 - p)^2 (1 - p + \frac{p^2}{2}) \\
    &= \rho_{01} \cdot \left[(1 - p)^2 (1 - p + \frac{p^2}{2})\right]^{n/2} 
\end{aligned},
\label{eq:noisycnotslinkingtwostepoffdiagonal}
\end{equation}

\bbbsection{Noisy CNOTs -- Repeater Protocol}
Identically to the previous case, we also recover the fact that every contribution from the diagonal terms which do not belong to the GHZ state, are always at least $\order{\epsilon^2}$. Using this fact, and calculating recursively from Eq. 19 in the main text, we get that:
\begin{equation}
\begin{aligned}
    \rho_{00}' = \rho_{11}' &= \rho_{00} \cdot \prod_{j \in D(n)} \tilde{h}(p) + \order{\epsilon^2} \\
    \tilde{h}(p) &= \left(1-p\right)^2 + \frac{p}{2} \left(1-\frac{p}{2}\right)
\end{aligned},
\label{eq:noisycnotslinkingrepeater}
\end{equation}
where $p = p_n$ is again the CNOT error, and we denote by $D(n)$ the set of nodes in the chain $L(n)$ which perform a deterministic CNOT. As for the off-diagonal terms, the result is again very simple, but slightly different:
\begin{equation}
\begin{aligned}
    \rho_{01}' = \tilde{\rho}_{10} &= \rho_{01} \cdot \prod_{j \in D(n)} (1 - p)^3 (1 - \frac{p}{2}) \\
\end{aligned}.
\label{eq:noisycnotslinkingrepeateroffdiagonal}
\end{equation}

\bbsection{Final Decoherence in Memories}

Finally, after each linking step is performed, the qubits stay in the memory suffering decoherence, which can be described by dephasing and amplitude-damping channels. Moreover, this step is identical to both protocols.

The dephasing contributions are the simplest to calculate. They contribute in the exact same manner as previously, affecting only the off-diagonal entries of the density matrix. The amplitude-damping contributions to this off-diagonal terms are also straightforward to calculate:

\begin{equation}
\begin{aligned}
    \rho_{01}'' = \rho_{10}'' = \rho_{01}' \prod_{i \in L(n)} \epsilon(t_i / T_2^{\mathrm{e,n}}) \sqrt{\epsilon(t_i / T_1^{\mathrm{e,n}})} \\
\end{aligned}.
\label{eq:finaldecoherencediagonal}
\end{equation}
where $T_{1,2}^{\mathrm{e,n}}$ is chosen accordingly, meaning, if the qubit is an electronic spin or a nuclear spin. If the protocol is the first two-step scheme, then it is always a nuclear spin. In the case of the hybrid scheme, one can have both a nuclear or electronic spin.

As for the diagonal terms, an approximation is made, similarly to the main text, in order to not calculate an exponential number of terms coming from each of the diagonal entries of the density matrix. Take that, when talking about the density matrix entries, except the GHZ entries (the first and the last, $\rho_{00}$ and $\rho_{11}$), the terms can be described by:

\begin{equation}
\begin{aligned}
    \rho_{ii} \mapsto \rho_{ii}' = \rho_{ii} + \order{\epsilon^2}
\end{aligned},
\end{equation}
since $\rho_{ii}$ is already $\order{\epsilon^1}$ for these entries. Each amplitude-damping channel applied to a qubit will make all the entries where such qubit takes value one, climb up the diagonal multiplied by a factor in $\order{\epsilon}$. For this reason, including a contribution from this error, we verify that the contributions come in at order at least $\order{\epsilon^2}$. Nonetheless, since there are many contributions in $\order{\epsilon^2}$, as there are qubits, and they might add-up to something not in $\order{\epsilon^2}$. This is the reason we also include them in the calculations. Let us first demonstrate an example of calculating diagonal terms of a 4-GHZ state, built from 3 different entangled pairs with $\{\mu_i,\nu_i\}_{i=1,2,3}$. 

\begin{equation}
\begin{aligned}
\rho_{00} =& (000) + \\
&+ \epsilon_1 (100) + \epsilon_2 (110) + \epsilon_3 (011) + \epsilon_4 (001) + \\
&+\epsilon_1 \epsilon_2 (010) + \epsilon_1 \epsilon_3 (111) + \epsilon_1 \epsilon_4 (101) + \epsilon_2\epsilon_3 (101) + \epsilon_2 \epsilon_4 (111) + \epsilon_3 \epsilon_4 (010) + \\
&+ \epsilon_1\epsilon_2\epsilon_3 (001) + \epsilon_1 \epsilon_2 \epsilon_4 (011) + \epsilon_1 \epsilon_3 \epsilon_4 (110) + \epsilon_2 \epsilon_3 \epsilon_4 (100) + \\
&+\epsilon_1 \epsilon_2 \epsilon_3 \epsilon_4 (000)
\end{aligned},
\label{eq:fullexpressionAD}
\end{equation}
where $(i_1 i_2 i_3) = \theta_{i_1}(\mu_1) \theta_{i_2}(\mu_2) \theta_{i_3}(\mu_3)$ and the $\theta$ function was defined in Eq. \ref{eq:cnotslinkingformula}, for example $(010) = (1-\mu_1) \mu_2 (1- \mu_3)$. Moreover, we abbreviate $\epsilon(t_i/T_1^{\mathrm{e,n}})$ as $\epsilon_i$, representing the error parameter from the amplitude-damping channel applied to qubit $i$. One can verify that in fact $(i_1 i_2 ... i_n) \sim O(\mu^{\sum_j i_j})$, and for the $k$th line in Eq. \ref{eq:cnotslinkingformula} there are $\binom{N}{k}$ terms (where $N$ is the number of qubits of the GHZ state). Given this, since the biggest terms in the $k$th line are of $\order{\epsilon^{k+1}}$, and there are at most $\binom{N}{k}$ of them, we can verify that our approximation is good if at least:

\begin{equation}
\begin{aligned}
\binom{N}{k}\cdot \epsilon^{k+1} &= N^{\underline{k}}\cdot \epsilon^{k+1} \\
&= N \cdot \frac{N-1}{1} \cdot \frac{N-2}{2} \cdots \frac{N-k}{k} \cdot \epsilon^{k+1} \\
&\leq \frac{(N\epsilon)^k}{k!}\cdot \epsilon \\
\implies \quad  N  \epsilon & \lesssim 1
\end{aligned}
\end{equation}

Since $T_1$ is not the shortcoming in current setups, we can expect that even for $t/T1 \sim  10 \mathrm{\mu s}/ 20 \mathrm{ms} $ we have $\epsilon \rightarrow 10^{-4}$ which means the approximations are valid for at least 10.000 qubits. Since the first line is the same as not adding any contributions from the final amplitude-damping, adding the second line results in the following expression for $\rho_{00}$:

\begin{equation}
\begin{aligned}
    \rho_{00}'' = \rho_{00}'+ \sum_{j =1 }^N (j) \epsilon_j 
\end{aligned},
\end{equation}
where $(j) = (0...01_{j-1}1_{j}0...0)$ for $1<j<N$, and with $(1) = (10...0)$ and $(N) = (0...01)$, where again, $(i_1 i_2 ... i_{N-1}) = \theta_{i_1}(\mu_1) \theta_{i_2}(\mu_2) ...  \theta_{i_{N-1}}(\mu_{N-1})$.
The only thing missing is calculating the last diagonal entry, which, from the form of the amplitude-damping channel applied to the entry $\ket{11...1}\bra{11...1}$ is simply:
\begin{equation}
\begin{aligned}
    \rho_{11}'' = \rho_{11}'\cdot \prod_{j =1 }^N (1-\epsilon_j) 
\end{aligned}.
\end{equation}

\bbsection{Randomized corrections}

GHZ states are highly symmetric states, derived directly from their form. Moreover, they belong to a set of states that can be generated using a limited amount of Clifford Gates. Among its implications, there is one which allows us to define a state solely from its set of stabilizers, as it is known that there is an equivalence between the set of stabilizer states and the ones that can be reached only using Clifford gates.

Denote the Pauli group on $n$ qubits by $\mathcal{G}_n$, we call $\ket{\psi}$ a stabilizer state if we can find a subset $S \subset \mathcal{G}_n$, such that $|S|=2^n$ and $\forall \ A \in S: A\ket{\psi} = \ket{\psi}$. In particular, the fidelity of a quantum state is invariant under any stabilizer. Taking this and a channel for which we choose the Kraus operators from the set of stabilizers (apart from a constant), we have that:
\begin{equation}
\begin{aligned}
    \mathcal{S} = \{\sqrt{p_i} A_i\}_{A_i \in S} , \text{ such that } \sum_i p_i = 1  , \text{ then: }\\
    F(\mathcal{S}(\rho),\ket{\psi}\bra{\psi}) = \sum_i p_i F(A_i \rho A_i^\dagger,\ket{\psi}\bra{\psi}) = F(\rho,\ket{\psi}\bra{\psi})
\end{aligned}
\label{eq:randomized}
\end{equation}

The GHZ state in particular is a stabilizer state, with one of the stabilizers being $S_X = \bigotimes_i \hat{X}_i$. This operation in particular can be seen as a transposition along the anti-diagonal, which we shall denote by $T_{AD}(\cdot)$. Here, by anti-diagonal of a matrix, we mean the elements that range from $\ket{0}\bra{2^n}$ to $\ket{2^n}\bra{0}$, $i.e.$ from the upper-right corner to the down-left corner of the matrix. This is an important notion to have, as the GHZ state is symmetric not only over the diagonal, but also over the anti-diagonal, in the computational-basis. 

However, the action induced by the amplitude-damping channel breaks this symmetry, as expected. To try and compensate for this symmetry breaking, we introduce the concept of randomized corrections. Consider the simplest example of a GHZ state with two qubits, the $\phi^+$ Bell pair. When creating this state in our protocol, there is a sequence of measurements with posterior corrections that are made. In particular, we use this fact to randomize the type of corrections sent, utilizing the $S_X$ stabilizer equivalence of the Bell pair $\psi^+$ (which is the state previous to the correction):

\begin{equation}
\begin{aligned}
    \mathbb{1} \otimes \hat{X} \ket{\psi^+} &= ( \mathbb{1} \otimes \hat{X} ) \circ (\hat{X} \otimes \hat{X} ) \ket{\psi^+} \\
    &= \hat{X} \otimes \mathbb{1} \ket{\psi^+} \\
    &= \ket{\phi^+}
\end{aligned},
\end{equation}
we can send half of the time the correction for the first qubit and half of the time the correction for the second qubit. This will create an ensemble with the same fidelity (given by Eq. \ref{eq:randomized}), and now the same symmetry as before. Equivalent to applying a $\hat{X}\otimes \hat{X}$ half the times, we can think of this operation as:
\begin{equation}
\begin{aligned}
    \mathrm{Sym}_\mathrm{A} (\rho) := \frac{\rho + T_{\mathrm{AD}} (\rho)}{2}
\end{aligned},
\end{equation}
which is the corresponding of symmetrizing matrices over their diagonal, by adding their transpose:
\begin{equation}
\begin{aligned}
    \mathrm{Sym} (\rho) := \frac{\rho + \rho^\mathrm{T}}{2}
\end{aligned}.
\end{equation}

In all the calculations in the main text we have used this as means to simplify the expressions and symmetrize the matrices, without actually changing the final fidelity output.

\vspace{0.3cm}
\bbsection{Fidelity of Accessing the QRAM}

In this appendix, we will discuss why the fidelity for accessing the QRAM will essentially be affected by the several GHZ states existent in each of its layers.

In terms of noisy operations, the operations that are used to create the GHZ states within each layer grow linearly with the number of memory cells. The number of operations used to route the bus qubit along the QRAM itself only grows logarithmically, as this is one of the main benefits from this architecture.

In terms of actual time the qubits are spent decohering on the memory, after the GHZ state is distributed within each layer, it takes $\order{\log N}$ for the bus qubit to be routed through the memory cells, which is at its worse, as big as the time it took to distribute the GHZ states. However, if these operations are photonic, this time is much smaller than that of creating the GHZ states as most of the decoherence occurs when creating the GHZ states.

Taking this into account, then the bus qubit will be routed via the noisy GHZ states. Trying to find a decent measure for the fidelity of access implies finding an average over all possible states for the address, execute the protocol, and measure the fidelity between the output state of the address plus bus qubit and the state for the case of a noiseless access. Even with a large representation of address states, this does not seem like a smart approach. If one looks at the fidelity of each of the GHZ states, it will essentially dictate how much of the bus qubit will be routed correctly, for any address state. Then, if each layer routes at a different level, the probability of a bus qubit being routed correctly along the tree, will be given by the product of being routed correctly at each of the layers. This happens since, if an error happens on one of the first layers, the bus qubit can never be routed to its correct memory cell. Then, this introduces an error on the final state. This is the reason why simply multiplying, for each layer, the fidelities of the GHZ states (as shown in the main text) translates in a good measure for the fidelity of accessing the QRAM.

\clearpage

\bsection{Supplementary Table}
In the following table, we present the physical values of important parameters that define the experimental setup, with the values reported reported in the main text and over Ref.~\cite{Chen2021}.

\begin{table}[H]
\centering
\begin{tabular}{c|c}
\hline \hline
\textbf{Single Qubit Gate Time} & 32~$\mathrm{ns}$ \\ \hline
\textbf{Qubit initialization time (${\ket{0}}$)} & 5~$\mu\mathrm{s}$ \\ \hline
\textbf{Nuclear CNOT Time} & 16~$\mu\mathrm{s}$ \\ \hline
\textbf{Electronic CNOT Time} & 29~$\mathrm{ns}$ \\ \hline
\textbf{Photon-Spin Interaction Time} & 0.1~$\mathrm{ns}$ \\ \hline
\textbf{Light Velocity in Medium} & $2\times 10^8$ \\ \hline
\textbf{Distance between Cavities} & $10~\mu\mathrm{m}$ \\ \hline \hline
\textbf{Electronic Qubits Damping Time}, $T_1$ & $\left[ 20~\mathrm{ms} , 200~\mathrm{ms} , 2~\mathrm{s} \right]$ \\ \hline
\textbf{Electronic Qubits Dephasing Time}, $T_2$ & $\left[ 10~\mathrm{ms} , 100~\mathrm{ms} , 1~\mathrm{s} \right]$ \\ \hline
\textbf{Nuclear Qubits Damping Time, $T_1$} & $\left[ 2~\mathrm{s} , 20~\mathrm{s} , 200~\mathrm{s} \right]$ \\ \hline
\textbf{Nuclear Qubits Dephasing Time, $T_2$} & $\left[ 1~\mathrm{s} , 10~\mathrm{s} , 100~\mathrm{s} \right]$ \\ \hline
\textbf{Electronic CNOT Error} & $\left[ 0, 10^{-5} , 10^{-4} , 10^{-3} , 10^{-2} \right]$ \\ \hline
\textbf{Nuclear CNOT Error} & $\left[ 0, 10^{-5} , 10^{-4} , 10^{-3} , 10^{-2} \right]$ \\ \hline
\textbf{Number of Simulations }& 100 \\ \hline
\end{tabular}
\label{tab:parameters}
\caption{Physical parameters used in simulations.}
\end{table}

\clearpage

\bsection{Supplementary Figures}

\begin{figure}[H]
\centering
\subfloat{\includegraphics[width=\linewidth]{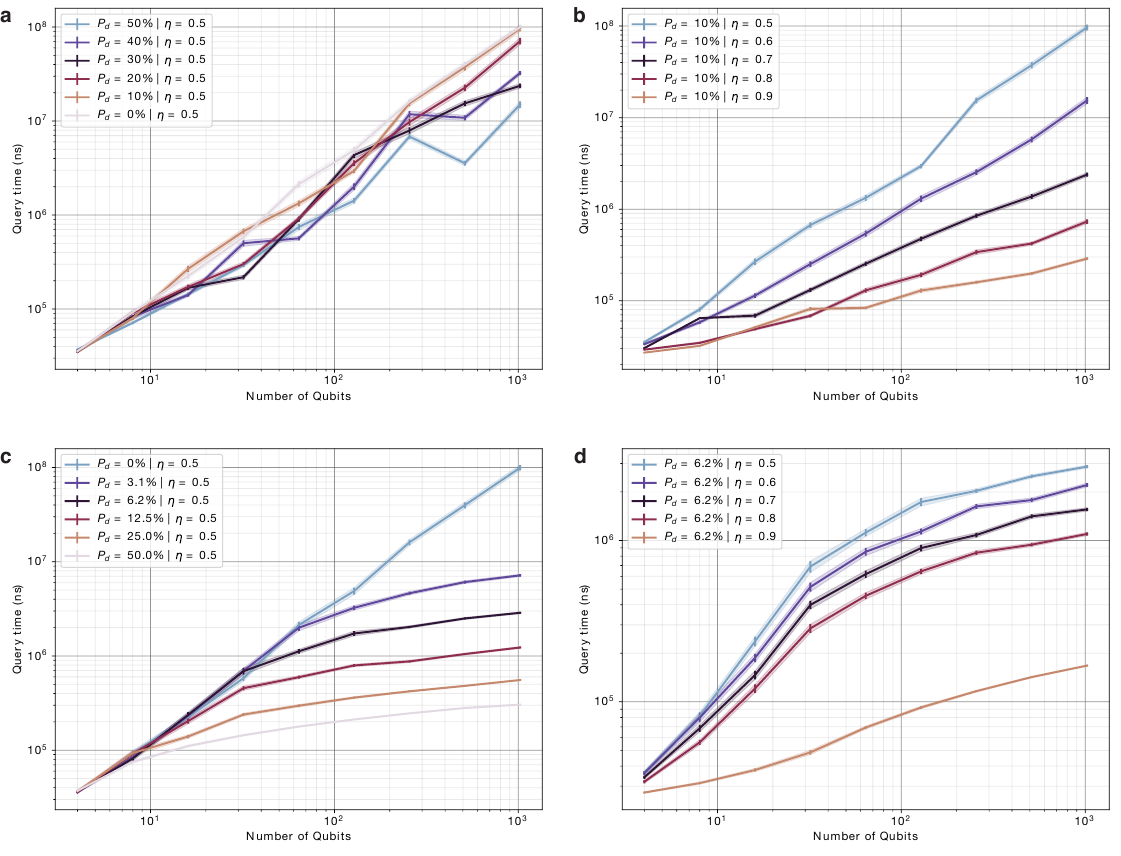}}
\caption{Query times for both types of placements of the deterministic linking steps under the repeater protocol, under several values of efficiency and amount of deterministic nodes. In \emph{(a)} and \emph{(b)}, we consider a random placement of deterministic nodes across each chain, and go from $0\%$ of deterministic nodes to $50\%$, and also vary the efficiency of the distributed CNOTs $\eta$, from $0.5$ to $0.9$. In \emph{(c)} and \emph{(d)}, we consider placing deterministic nodes only in the nodes used at the higher level time steps of the linking binary tree, in each chain. We vary the first deterministic layer from the 3rd to the 6th ($\log_2(N)-D \in \{2,3,4,5\}$, see Fig.5 in the main text), meaning that the probability of being a deterministic node in each layer of the QRAM is approximately $P_\mathrm{d} \approx 2^{-(\log_2(N)-D)}$. We also vary the efficiency of the distributed CNOTs $\eta$, from $0.5$ to $0.9$ as before.  }
\label{fig:querytimes_repeater_full}
\end{figure}

\clearpage

\begin{figure}[H]
\centering
\subfloat{\includegraphics[width=\linewidth]{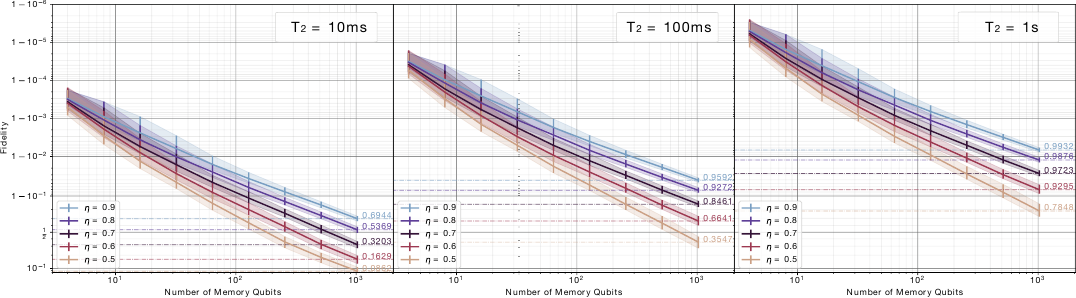}}
\caption{Fidelity scaling for different values of the dephasing time $T_2$, ranging from $10 \mathrm{ms}$ to $1 \mathrm{s}$ (with $T_1$ fixed at $2~\mathrm{s}$). The simulations are for the completely probabilistic execution of the linking step ($P_\mathrm{d} = 0\%$), meaning there are no deterministic CNOTs being executed to create the GHZ states within each layer of the QRAM. We present different simulations for several possible values for the efficiency of each distributed CNOT ($i.e.$ the probability of success of each of the distributed CNOT), namely $\eta \in \{0.5, 0.6, 0.7, 0.8, 0.9\}$. }
\label{fig:fidelities_1}
\end{figure} 

In Supplementary Figures~\ref{fig:fidelities_2}, \ref{fig:fidelities_3} and \ref{fig:fidelities_4} we present combinations of error parameters (increasing $T_2$ times across the horizontal) and amounts of deterministic nodes (decreasing $P_\mathrm{d}$ across the vertical) for CNOT errors in the order of $1\%$, $0.1\%$ and $0.01\%$, respectively. For each of the plots, we also plot, for comparison purposes with equivalent error parameters, the initial TD-QRAM architecture, to verify when there is an advantage to use the hybrid scheme.

\vfill
\pagebreak

\begin{figure}[H]
\centering
\subfloat{\includegraphics[width=\linewidth]{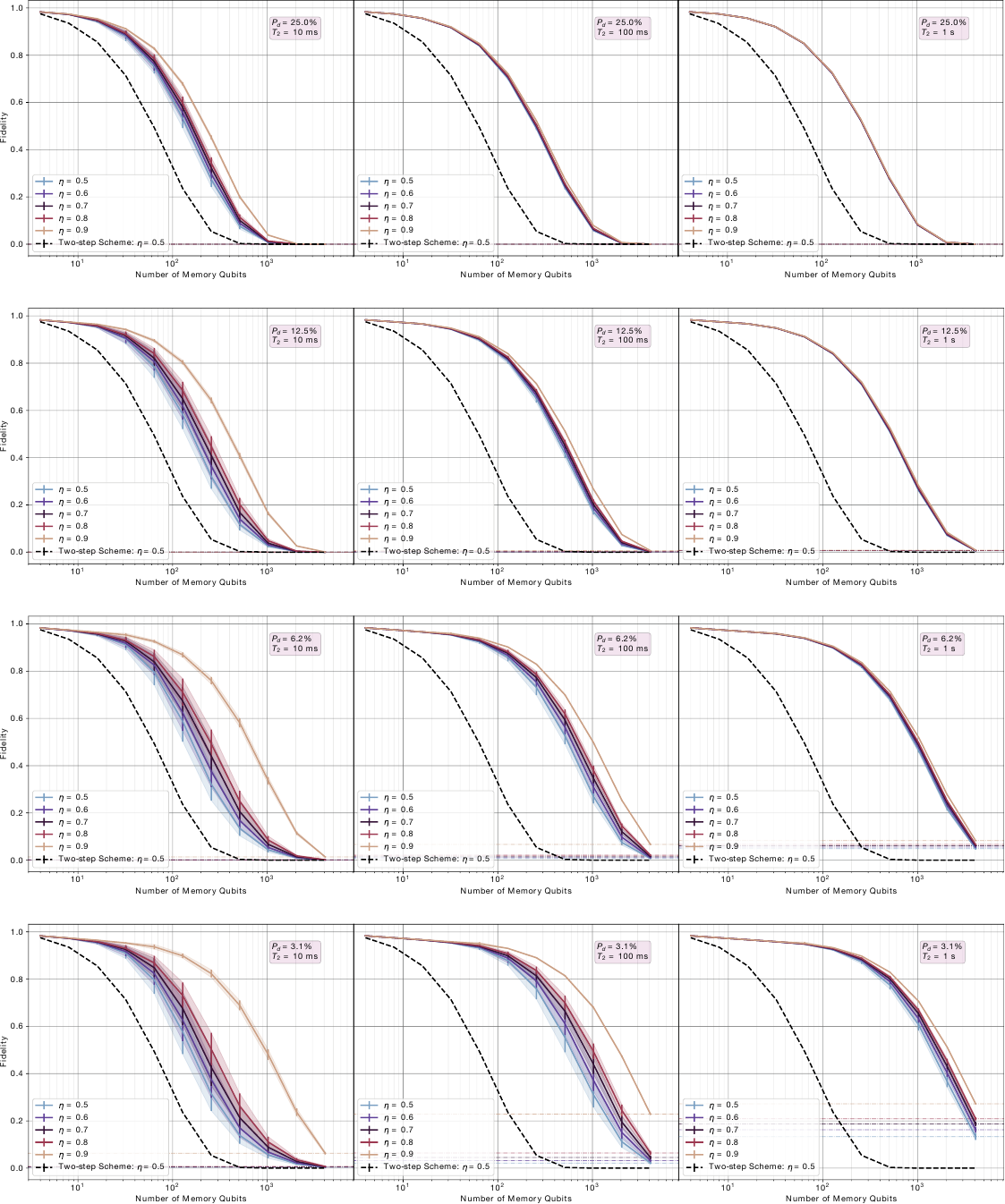}} 
\caption{Fidelity scaling in function of the number of memory qubits for a TS-QRAM. Dephasing time $T_2$ ranging from $10 \mathrm{ms}$ to $1 \mathrm{s}$. Amplitude-damping time $T_1$ fixed at $2~\mathrm{s}$. Deterministic CNOT error probability fixed at $1\%$. The simulations are for the hybrid scheme under different possible amounts of deterministic nodes ($P_\mathrm{d} = \{3.1\% , 6.2\%, 12.5\%, 25\% \}$). We present different simulations for several possible values for the efficiency of each distributed CNOT ($i.e.$ the probability of success of each of the distributed CNOT), namely $\eta \in \{0.5, 0.6, 0.7, 0.8, 0.9\}$. }
\label{fig:fidelities_2}
\end{figure} 

\begin{figure}[H]
\centering
\subfloat{\includegraphics[width=\linewidth]{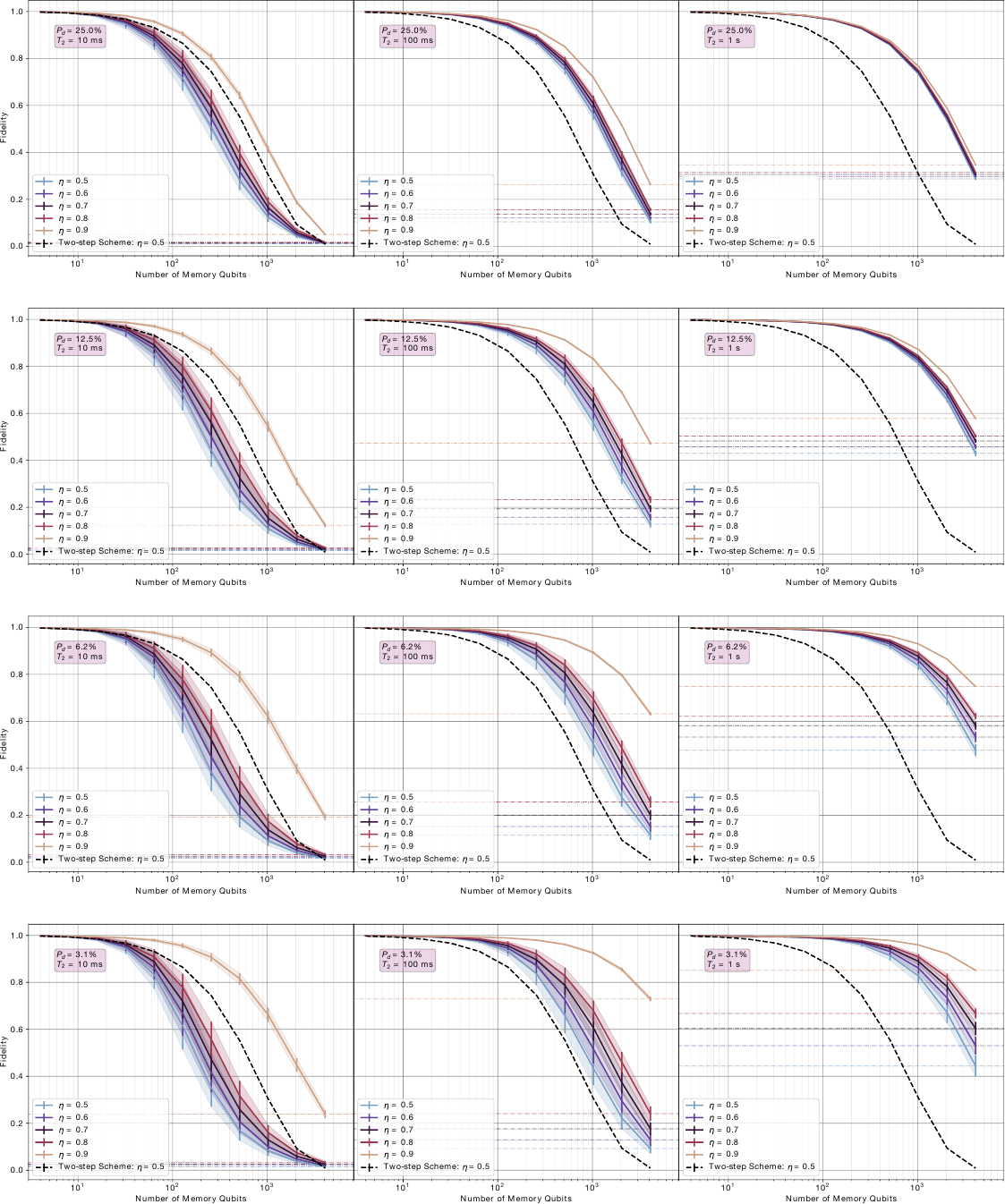}}
\caption{Fidelity scaling in function of the number of memory qubits for a TS-QRAM. Dephasing time $T_2$ ranging from $10 \mathrm{ms}$ to $1 \mathrm{s}$. Amplitude-damping time $T_1$ fixed at $2~\mathrm{s}$. Deterministic CNOT error probability fixed at $0.1\%$. The simulations are for the hybrid scheme under different possible amounts of deterministic nodes ($P_\mathrm{d} = \{3.1\% , 6.2\%, 12.5\%, 25\% \}$). We present different simulations for several possible values for the efficiency of each distributed CNOT ($i.e.$ the probability of success of each of the distributed CNOT), namely $\eta \in \{0.5, 0.6, 0.7, 0.8, 0.9\}$. }
\label{fig:fidelities_3}
\end{figure}  

\begin{figure}[H]
\centering
\subfloat{\includegraphics[width=\linewidth]{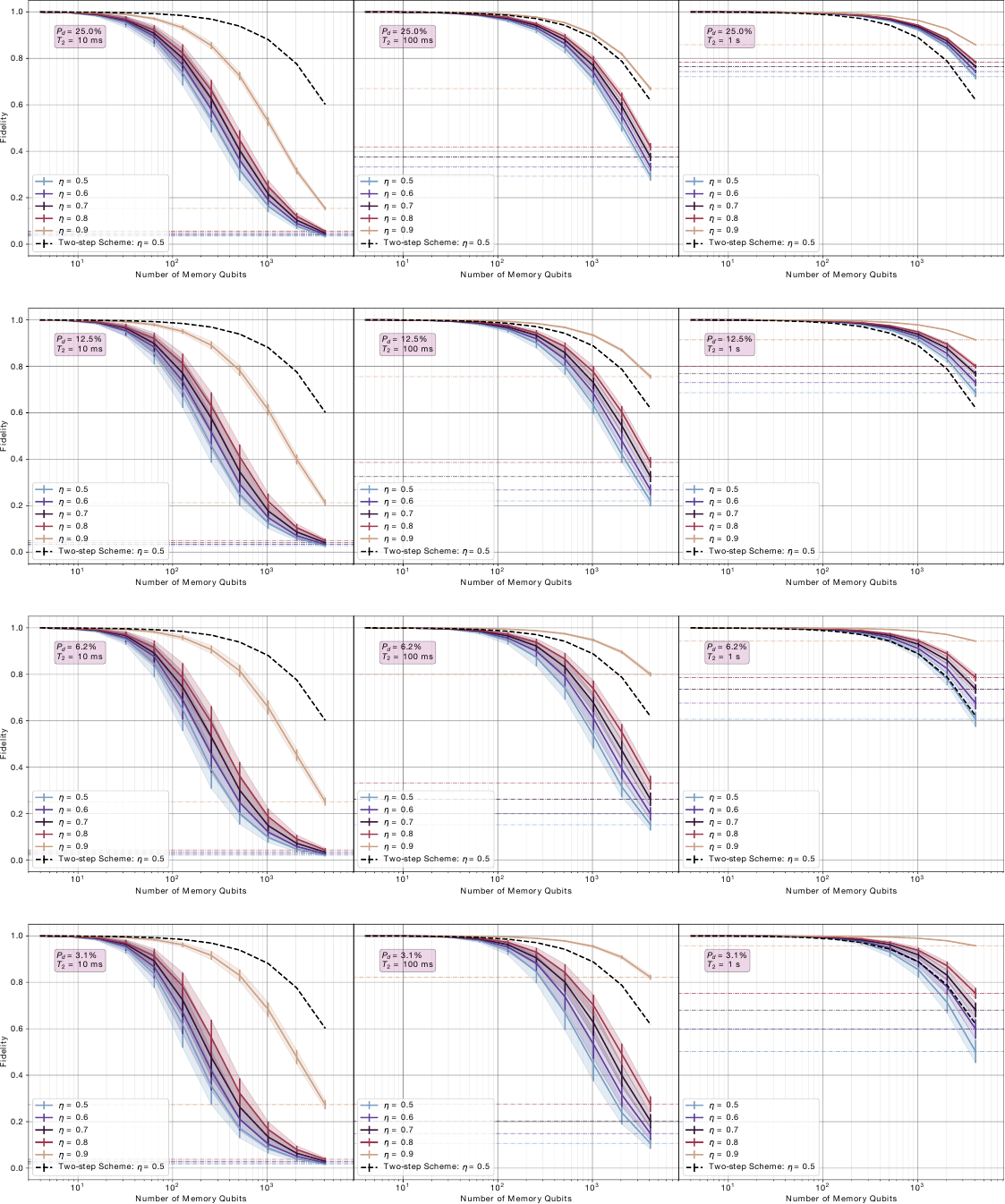}}
\caption{Fidelity scaling in function of the number of memory qubits for a TS-QRAM. Dephasing time $T_2$ ranging from $10 \mathrm{ms}$ to $1 \mathrm{s}$. Amplitude-damping time $T_1$ fixed at $2~\mathrm{s}$. Deterministic CNOT error probability fixed at $0.01\%$. The simulations are for the hybrid scheme under different possible amounts of deterministic nodes ($P_\mathrm{d} = \{3.1\% , 6.2\%, 12.5\%, 25\% \}$). We present different simulations for several possible values for the efficiency of each distributed CNOT ($i.e.$ the probability of success of each of the distributed CNOT), namely $\eta \in \{0.5, 0.6, 0.7, 0.8, 0.9\}$. }
\label{fig:fidelities_4}
\end{figure}


\begin{thebibliography}{10}
\expandafter\ifx\csname url\endcsname\relax
  \def\url#1{\texttt{#1}}\fi
\expandafter\ifx\csname urlprefix\endcsname\relax\def\urlprefix{URL }\fi
\providecommand{\bibinfo}[2]{#2}
\providecommand{\eprint}[2][]{\url{#2}}

\bibitem{Biamonte2017}
\bibinfo{author}{Biamonte, J.} \emph{et~al.}
\newblock \bibinfo{title}{Quantum machine learning}.
\newblock \emph{\bibinfo{journal}{Nature}} \textbf{\bibinfo{volume}{549}},
  \bibinfo{pages}{195--202} (\bibinfo{year}{2017}).

\bibitem{Harrow2009}
\bibinfo{author}{Harrow, A.~W.}, \bibinfo{author}{Hassidim, A.} \&
  \bibinfo{author}{Lloyd, S.}
\newblock \bibinfo{title}{Quantum {{Algorithm}} for {{Linear Systems}} of
  {{Equations}}}.
\newblock \emph{\bibinfo{journal}{Phys. Rev. Lett.}}
  \textbf{\bibinfo{volume}{103}}, \bibinfo{pages}{150502}
  (\bibinfo{year}{2009}).

\bibitem{Kiani2020}
\bibinfo{author}{Kiani, B.~T.}, \bibinfo{author}{Villanyi, A.} \&
  \bibinfo{author}{Lloyd, S.}
\newblock \bibinfo{title}{Quantum {{Medical Imaging Algorithms}}}.
\newblock \eprint{Pre-print available at \href{https://doi.org/10.48550/arXiv.2004.02036}{https://arxiv.org/abs/2004.02036}} (\bibinfo{year}{2020}).

\bibitem{Grover1996}
\bibinfo{author}{Grover, L.~K.}
\newblock \bibinfo{title}{A fast quantum mechanical algorithm for database
  search}.
\newblock In \emph{\bibinfo{booktitle}{Proceedings of the Twenty-Eighth Annual
  {{ACM}} Symposium on {{Theory}} of {{Computing}}}}, {{STOC}} '96,
  \bibinfo{pages}{212--219} (\bibinfo{publisher}{{Association for Computing
  Machinery}}, \bibinfo{address}{{New York, NY, USA}}, \bibinfo{year}{1996}).

\bibitem{Chen2021}
\bibinfo{author}{Chen, K.~C.}, \bibinfo{author}{Dai, W.},
  \bibinfo{author}{{Errando-Herranz}, C.}, \bibinfo{author}{Lloyd, S.} \&
  \bibinfo{author}{Englund, D.}
\newblock \bibinfo{title}{Scalable and {{High-Fidelity Quantum Random Access
  Memory}} in {{Spin-Photon Networks}}}.
\newblock \emph{\bibinfo{journal}{PRX Quantum}} \textbf{\bibinfo{volume}{2}},
  \bibinfo{pages}{030319} (\bibinfo{year}{2021}).

\bibitem{Jaeger1997}
\bibinfo{author}{Jaeger, R.} \& \bibinfo{author}{Blalock, T.}
\newblock \emph{\bibinfo{title}{Microelectronic {{Circuit Design}}}}
  (\bibinfo{publisher}{{McGraw-Hill Education}}, \bibinfo{address}{{New York,
  NY}}, \bibinfo{year}{1997}), \bibinfo{edition}{4th edition} edn.

\bibitem{Giovannetti2008}
\bibinfo{author}{Giovannetti, V.}, \bibinfo{author}{Lloyd, S.} \&
  \bibinfo{author}{MacCone, L.}
\newblock \bibinfo{title}{Architectures for a quantum random access memory}.
\newblock \emph{\bibinfo{journal}{Phys. Rev. A}} \textbf{\bibinfo{volume}{78}}, \bibinfo{pages}{052310}
  (\bibinfo{year}{2008}).

\bibitem{Hann2021a}
\bibinfo{author}{Hann, C.~T.}, \bibinfo{author}{Lee, G.},
  \bibinfo{author}{Girvin, S.} \& \bibinfo{author}{Jiang, L.}
\newblock \bibinfo{title}{Resilience of {{Quantum Random Access Memory}} to
  {{Generic Noise}}}.
\newblock \emph{\bibinfo{journal}{PRX Quantum}} \textbf{\bibinfo{volume}{2}},
  \bibinfo{pages}{020311} (\bibinfo{year}{2021}).

\bibitem{Giovannetti2008a}
\bibinfo{author}{Giovannetti, V.}, \bibinfo{author}{Lloyd, S.} \&
  \bibinfo{author}{MacCone, L.}
\newblock \bibinfo{title}{Quantum random access memory}.
\newblock \emph{\bibinfo{journal}{Phys. Rev. Lett.}}
  \textbf{\bibinfo{volume}{100}}, \bibinfo{pages}{160501}
  (\bibinfo{year}{2008}).

\bibitem{Toth2012a}
\bibinfo{author}{T{\'o}th, G.}
\newblock \bibinfo{title}{Multipartite entanglement and high-precision
  metrology}.
\newblock \emph{\bibinfo{journal}{Phys. Rev. A}}
  \textbf{\bibinfo{volume}{85}}, \bibinfo{pages}{022322}
  (\bibinfo{year}{2012}).

\bibitem{Sidhu2019}
\bibinfo{author}{Sidhu, J.~S.} \& \bibinfo{author}{Kok, P.}
\newblock \bibinfo{title}{A {{Geometric Perspective}} on {{Quantum Parameter
  Estimation}}}.
\newblock \emph{\bibinfo{journal}{AVS Quantum Sci.}}
  \textbf{\bibinfo{volume}{2}}, \bibinfo{pages}{014701} (\bibinfo{year}{2019}).

\bibitem{Murta2020}
\bibinfo{author}{Murta, G.}, \bibinfo{author}{Grasselli, F.},
  \bibinfo{author}{Kampermann, H.} \& \bibinfo{author}{Bru{\ss}, D.}
\newblock \bibinfo{title}{Quantum {{Conference Key Agreement}}: {{A Review}}}.
\newblock \emph{\bibinfo{journal}{Adv. Quantum Technol.}}
  \textbf{\bibinfo{volume}{3}}, \bibinfo{pages}{2000025}
  (\bibinfo{year}{2020}).

\bibitem{Wehner2018}
\bibinfo{author}{Wehner, S.}, \bibinfo{author}{Elkouss, D.} \&
  \bibinfo{author}{Hanson, R.}
\newblock \bibinfo{title}{Quantum internet: {{A}} vision for the road ahead}.
\newblock \emph{\bibinfo{journal}{Science}} \textbf{\bibinfo{volume}{362}},
  \bibinfo{pages}{eaam9288} (\bibinfo{year}{2018}).

\bibitem{Alshowkan2021a}
\bibinfo{author}{Alshowkan, M.} \emph{et~al.}
\newblock \bibinfo{title}{Reconfigurable {{Quantum Local Area Network Over
  Deployed Fiber}}}.
\newblock \emph{\bibinfo{journal}{PRX Quantum}} \textbf{\bibinfo{volume}{2}},
  \bibinfo{pages}{040304} (\bibinfo{year}{2021}).

\bibitem{VandenNest2011}
\bibinfo{author}{{Van den Nest}, M.}
\newblock \bibinfo{title}{Simulating quantum computers with probabilistic
  methods}.
\newblock \emph{\bibinfo{journal}{Quantum Inf. Comput.}}
  \textbf{\bibinfo{volume}{11}}, \bibinfo{pages}{784--812}
  (\bibinfo{year}{2011}).

\bibitem{Jozsa2014}
\bibinfo{author}{Jozsa, R.} \& \bibinfo{author}{{van den Nest}, M.}
\newblock \bibinfo{title}{Classical simulation complexity of extended clifford
  circuits}.
\newblock \emph{\bibinfo{journal}{Quantum Inf. Comput.}}
  \textbf{\bibinfo{volume}{14}}, \bibinfo{pages}{633--648}
  (\bibinfo{year}{2014}).

\bibitem{Takahashi2020}
\bibinfo{author}{Takahashi, Y.}, \bibinfo{author}{Takeuchi, Y.} \&
  \bibinfo{author}{Tani, S.}
\newblock \bibinfo{title}{Classically {{Simulating Quantum Circuits}} with
  {{Local Depolarizing Noise}}}.
\newblock \emph{\bibinfo{journal}{Theor. Comput. Sci.}} \textbf{\bibinfo{volume}{893}}, \bibinfo{pages}{117--132}  (\bibinfo{year}{2021}).

\bibitem{Bhaskar2020}
\bibinfo{author}{Bhaskar, M.~K.} \emph{et~al.}
\newblock \bibinfo{title}{Experimental demonstration of memory-enhanced quantum
  communication}.
\newblock \emph{\bibinfo{journal}{Nature}} \textbf{\bibinfo{volume}{580}},
  \bibinfo{pages}{60--64} (\bibinfo{year}{2020}).

\bibitem{Wan2020}
\bibinfo{author}{Wan, N.~H.} \emph{et~al.}
\newblock \bibinfo{title}{Large-scale integration of artificial atoms in hybrid
  photonic circuits}.
\newblock \emph{\bibinfo{journal}{Nature}} \textbf{\bibinfo{volume}{583}},
  \bibinfo{pages}{226--231} (\bibinfo{year}{2020}).

\bibitem{Nguyen2019}
\bibinfo{author}{Nguyen, C.~T.} \emph{et~al.}
\newblock \bibinfo{title}{An integrated nanophotonic quantum register based on
  silicon-vacancy spins in diamond}.
\newblock \emph{\bibinfo{journal}{Phys. Rev. B}}
  \textbf{\bibinfo{volume}{100}}, \bibinfo{pages}{165428}
  (\bibinfo{year}{2019}).

\bibitem{Bradley2021}
\bibinfo{author}{Bradley, C.~E.} \emph{et~al.}
\newblock \bibinfo{title}{Robust quantum-network memory based on spin qubits in
  isotopically engineered diamond}.
\newblock \eprint{Pre-print available at \href{
https://doi.org/10.48550/arXiv.2111.09772}{https://arxiv.org/abs/2111.09772}} (\bibinfo{year}{2021}).

\bibitem{Chen2021a}
\bibinfo{author}{Chen, K.~C.}, \bibinfo{author}{Bersin, E.} \&
  \bibinfo{author}{Englund, D.}
\newblock \bibinfo{title}{A polarization encoded photon-to-spin interface}.
\newblock \emph{\bibinfo{journal}{npj Quantum Inf.}}
  \textbf{\bibinfo{volume}{7}}, \bibinfo{pages}{1--6} (\bibinfo{year}{2021}).

\bibitem{Sukachev2017}
\bibinfo{author}{Sukachev, D.~D.} \emph{et~al.}
\newblock \bibinfo{title}{Silicon-{{Vacancy Spin Qubit}} in {{Diamond}}: {{A
  Quantum Memory Exceeding}} 10 ms with {{Single-Shot State Readout}}}.
\newblock \emph{\bibinfo{journal}{Phys. Rev. Lett.}}
  \textbf{\bibinfo{volume}{119}}, \bibinfo{pages}{223602}
  (\bibinfo{year}{2017}).

\bibitem{Duan2004a}
\bibinfo{author}{Duan, L.-M.} \& \bibinfo{author}{Kimble, H.~J.}
\newblock \bibinfo{title}{A scheme for preparation of multi-atom entanglement
  by detecting the cavity decay and analysis of its implementation}.
\newblock In \emph{\bibinfo{booktitle}{Quantum {{Communications}} and {{Quantum
  Imaging}}}}, vol. \bibinfo{volume}{5161}, \bibinfo{pages}{40--47}
  (\bibinfo{publisher}{{SPIE}}, \bibinfo{year}{2004}).

\bibitem{Calderon-Vargas2019}
\bibinfo{author}{{Calderon-Vargas}, F.~A.} \emph{et~al.}
\newblock \bibinfo{title}{Fast high-fidelity entangling gates for spin qubits
  in {{Si}} double quantum dots}.
\newblock \emph{\bibinfo{journal}{Phys. Rev. B}}
  \textbf{\bibinfo{volume}{100}}, \bibinfo{pages}{035304}
  (\bibinfo{year}{2019}).

\bibitem{Coopmans2022}
\bibinfo{author}{Coopmans, T.}, \bibinfo{author}{Brand, S.} \&
  \bibinfo{author}{Elkouss, D.}
\newblock \bibinfo{title}{Improved analytical bounds on delivery times of
  long-distance entanglement}.
\newblock \emph{\bibinfo{journal}{Phys. Rev. A}}
  \textbf{\bibinfo{volume}{105}}, \bibinfo{pages}{012608}
  (\bibinfo{year}{2022}).

\bibitem{Duan2004}
\bibinfo{author}{Duan, L.-M.} \& \bibinfo{author}{Kimble, H.~J.}
\newblock \bibinfo{title}{Scalable {{Photonic Quantum Computation}} through
  {{Cavity-Assisted Interactions}}}.
\newblock \emph{\bibinfo{journal}{Phys. Rev. Lett.}}
  \textbf{\bibinfo{volume}{92}}, \bibinfo{pages}{127902}
  (\bibinfo{year}{2004}).

\bibitem{Coopmans2020e}
\bibinfo{author}{Coopmans, T.} \emph{et~al.}
\newblock \bibinfo{title}{{{NetSquid}}, a {{NETwork Simulator}} for {{QUantum
  Information}} using {{Discrete}} events}.
\newblock \emph{\bibinfo{journal}{Commun. Phys.}}
  \textbf{\bibinfo{volume}{4}}, \bibinfo{pages}{164} (\bibinfo{year}{2021}).

\bibitem{Dai2020}
\bibinfo{author}{Dai, W.}, \bibinfo{author}{Peng, T.} \& \bibinfo{author}{Win,
  M.~Z.}
\newblock \bibinfo{title}{Optimal {{Remote Entanglement Distribution}}}.
\newblock \emph{\bibinfo{journal}{IEEE J. Sel. Areas Commun.}} \textbf{\bibinfo{volume}{38}}, \bibinfo{pages}{540--556}
  (\bibinfo{year}{2020}).

\bibitem{Childress2006}
\bibinfo{author}{Childress, L.} \emph{et~al.}
\newblock \bibinfo{title}{Coherent {{Dynamics}} of {{Coupled Electron}} and
  {{Nuclear Spin Qubits}} in {{Diamond}}}.
\newblock \emph{\bibinfo{journal}{Science}} \textbf{\bibinfo{volume}{314}},
  \bibinfo{pages}{281--285} (\bibinfo{year}{2006}).

\bibitem{Findler2020}
\bibinfo{author}{Findler, C.}, \bibinfo{author}{Lang, J.},
  \bibinfo{author}{Osterkamp, C.}, \bibinfo{author}{Nesl{\'a}dek, M.} \&
  \bibinfo{author}{Jelezko, F.}
\newblock \bibinfo{title}{Indirect overgrowth as a synthesis route for superior
  diamond nano sensors}.
\newblock \emph{\bibinfo{journal}{Sci. Rep.}}
  \textbf{\bibinfo{volume}{10}}, \bibinfo{pages}{22404} (\bibinfo{year}{2020}).

\bibitem{Brand2020}
\bibinfo{author}{Brand, S.}, \bibinfo{author}{Coopmans, T.} \&
  \bibinfo{author}{Elkouss, D.}
\newblock \bibinfo{title}{Efficient {{Computation}} of the {{Waiting Time}} and
  {{Fidelity}} in {{Quantum Repeater Chains}}}.
\newblock \emph{\bibinfo{journal}{IEEE J. Sel. Areas Commun.}} \textbf{\bibinfo{volume}{38}}, \bibinfo{pages}{619--639}
  (\bibinfo{year}{2020}).

\bibitem{Pichler2017}
\bibinfo{author}{Pichler, H.}, \bibinfo{author}{Choi, S.},
  \bibinfo{author}{Zoller, P.} \& \bibinfo{author}{Lukin, M.~D.}
\newblock \bibinfo{title}{Universal photonic quantum computation via
  time-delayed feedback}.
\newblock \emph{\bibinfo{journal}{Proceedings of the National Academy of
  Sciences}} \textbf{\bibinfo{volume}{114}}, \bibinfo{pages}{11362--11367}
  (\bibinfo{year}{2017}).

\bibitem{Larsen2019}
\bibinfo{author}{Larsen, M.~V.}, \bibinfo{author}{Guo, X.},
  \bibinfo{author}{Breum, C.~R.}, \bibinfo{author}{{Neergaard-Nielsen}, J.~S.}
  \& \bibinfo{author}{Andersen, U.~L.}
\newblock \bibinfo{title}{Deterministic generation of a two-dimensional cluster
  state}.
\newblock \emph{\bibinfo{journal}{Science}} \textbf{\bibinfo{volume}{366}},
  \bibinfo{pages}{369--372} (\bibinfo{year}{2019}).

\bibitem{Russo2019}
\bibinfo{author}{Russo, A.}, \bibinfo{author}{Barnes, E.} \&
  \bibinfo{author}{Economou, S.~E.}
\newblock \bibinfo{title}{Generation of arbitrary all-photonic graph states
  from quantum emitters}.
\newblock \emph{\bibinfo{journal}{New J. Phys.}}
  \textbf{\bibinfo{volume}{21}}, \bibinfo{pages}{055002}
  (\bibinfo{year}{2019}).

\bibitem{Pant2019a}
\bibinfo{author}{Pant, M.}, \bibinfo{author}{Towsley, D.},
  \bibinfo{author}{Englund, D.} \& \bibinfo{author}{Guha, S.}
\newblock \bibinfo{title}{Percolation thresholds for photonic quantum
  computing}.
\newblock \emph{\bibinfo{journal}{Nat. Commun.}}
  \textbf{\bibinfo{volume}{10}}, \bibinfo{pages}{1070} (\bibinfo{year}{2019}).

\bibitem{Uppu2020}
\bibinfo{author}{Uppu, R.} \emph{et~al.}
\newblock \bibinfo{title}{Scalable integrated single-photon source}.
\newblock \emph{\bibinfo{journal}{Sci. Adv.}}
  \textbf{\bibinfo{volume}{6}}, \bibinfo{pages}{eabc8268}
  (\bibinfo{year}{2020}).

\bibitem{Michaels2021}
\bibinfo{author}{Michaels, C.~P.} \emph{et~al.}
\newblock \bibinfo{title}{Multidimensional cluster states using a single
  spin-photon interface coupled strongly to an intrinsic nuclear register}.
\newblock \emph{\bibinfo{journal}{Quantum}} \textbf{\bibinfo{volume}{5}},
  \bibinfo{pages}{565} (\bibinfo{year}{2021}).

\bibitem{Nickerson2014}
\bibinfo{author}{Nickerson, N.~H.}, \bibinfo{author}{Fitzsimons, J.~F.} \&
  \bibinfo{author}{Benjamin, S.~C.}
\newblock \bibinfo{title}{Freely {{Scalable Quantum Technologies Using Cells}}
  of 5-to-50 {{Qubits}} with {{Very Lossy}} and {{Noisy Photonic Links}}}.
\newblock \emph{\bibinfo{journal}{Phys. Rev. X}}
  \textbf{\bibinfo{volume}{4}}, \bibinfo{pages}{041041} (\bibinfo{year}{2014}).

\bibitem{Nemoto2014}
\bibinfo{author}{Nemoto, K.} \emph{et~al.}
\newblock \bibinfo{title}{Photonic {{Architecture}} for {{Scalable Quantum
  Information Processing}} in {{Diamond}}}.
\newblock \emph{\bibinfo{journal}{Phys. Rev. X}}
  \textbf{\bibinfo{volume}{4}}, \bibinfo{pages}{031022} (\bibinfo{year}{2014}).

\bibitem{Choi2019}
\bibinfo{author}{Choi, H.}, \bibinfo{author}{Pant, M.}, \bibinfo{author}{Guha,
  S.} \& \bibinfo{author}{Englund, D.}
\newblock \bibinfo{title}{Percolation-based architecture for cluster state
  creation using photon-mediated entanglement between atomic memories}.
\newblock \emph{\bibinfo{journal}{npj Quantum Inf.}}
  \textbf{\bibinfo{volume}{5}}, \bibinfo{pages}{104} (\bibinfo{year}{2019}).

\end{thebibliography}
\end{document}